\documentclass[12pt]{article}
\usepackage[dvips]{graphicx}
\usepackage{amsmath}
\usepackage{graphicx}
\usepackage{amsfonts}
\usepackage{amssymb}
\usepackage{epsfig}
\usepackage{subfigure}
\usepackage[usenames]{color}

\newcommand{\be}{\begin{equation}}
\newcommand{\ee}{\end{equation}}
\newcommand{\bea}{\begin{eqnarray}}
\newcommand{\eea}{\end{eqnarray}}
\newcommand{\ba}{\begin{array}}
\newcommand{\ea}{\end{array}}
\newcommand{\beas}{\begin{eqnarray*}}
\newcommand{\eeas}{\end{eqnarray*}}
\newcommand{\bes}{\begin{equation*}}
\newcommand{\ees}{\end{equation*}}

\def\i2           {\mbox{$\frac{i}{2}$}}

\begin{document}
\title{\bf Magnetic properties of holographic multiquarks in the quark-gluon plasma}

\author{Piyabut Burikham\thanks{Email:piyabut@gmail.com, piyabut.b@chula.ac.th}\\
{\small {\em Theoretical High-Energy Physics and Cosmology Group,
Department of Physics,}}\\
{\small {\em Faculty of Science, Chulalongkorn University, Bangkok
10330, Thailand}}}

\maketitle

\begin{abstract}
\noindent  \\

We study the magnetic properties of the coloured multiquark states
in the quark-gluon plasma where the gluons are deconfined and the
chiral symmetry is still broken, using the Sakai-Sugimoto model.
There are two possible magnetized multiquark configurations. Both
configurations converge to the same configuration at the critical
field and temperature before they dissociate altogether either
into less coloured multiquarks or into other phases for a fixed
density. It is also found that the multiquarks with higher colour
charges respond more to the external magnetic field in both the
magnetization and the degree of chiral symmetry breaking. Magnetic
field also makes it more difficult for multiquark states with
large colour charges to satisfy the equilibrium condition of the
configuration in the gravity dual picture.  As long as the
chemical potential $\mu
> \mu_{onset}$, the magnetized multiquarks phase is
thermodynamically preferred over the magnetized vacuum.  Pure pion
gradient and the chiral-symmetric quark-gluon
plasma~($\chi_S$-QGP) phase for the general Sakai-Sugimoto model
are discussed and compared with the multiquark phase in the
presence of the magnetic field.  It is found that at large
densities and moderate fields, the mixed phase of multiquarks and
the pion gradient is thermodynamically preferred over the
$\chi_S$-QGP.

\end{abstract}

\newpage
\section{Introduction}

There has been increasing interest in the study of nuclear phase
structure as well as properties of a number of nuclear phases,
especially the quark-gluon plasma in the recent few years.  This
is due to the new perspective in the nature of strongly
interacting gauge theory from the holographic principle. Motivated
by the AdS/CFT correspondence~\cite{maldacena,agmoo}, a number of
gravity models was constructed to provide shadow gauge theories
which share certain essential features with the QCD in the strong
coupling regime.  Sakai and Sugimoto~\cite{ss lowE,ss more}
proposed a toy holographic model of QCD where chiral symmetry
breaking can be addressed.  In Sakai-Sugimoto model, gluon
deconfinement and chiral symmetry restoration are two distinct
phase transitions.  For non-antipodal case, the chiral symmetry
restoration occurs at higher temperature than the gluon
deconfinement~\cite{asy}, therefore it is possible to have a
nuclear phase where gluons are deconfined while the quarks and
antiquarks could still form colour bound states.

Bergman, Lifschytz, and Lippert~\cite{bll} shows that when the
baryon density is sufficiently large and the temperature is not
too high, gluon-deconfined phase with broken chiral symmetry
accommodates a nuclear phase where baryons can exist with
thermodynamical stability.  Even though the baryons can exist
within the phase, the quark matters containing only free quarks or
antiquarks do not share the same thermodynamical stability.  This
can be understood as a sign of chiral symmetry breaking, the
quarks prefer to be bound together by gluon exchanges in this
highly-densed thermal soup.  Interestingly, further investigations
into whether colour multiquark states in general could exist
within this nuclear phase give positive results~\cite{bch}.

It was suggested quite a while ago in Ref.~\cite{BISY} that it is
possible to have $k<N$-baryons in $\mathcal{N}_{SUSY}=4$
background.  In the gluon-deconfined phase, since free strings
solution is allowed in the corresponding gravity dual
theories~\cite{abl}, the coloured states could also exist in the
plasma. Various possibilities of exotic multiquark states are
studied in Ref.~\cite{gint}-\cite{Wen}.  Colour multiquark states
in the gluon-deconfined plasma are studied in Ref.~\cite{bch}
where $k>N$-baryons as well as other classes of exotic multiquark
states including $N+\bar{k}$-baryons and bound state of $j$ mesons
are investigated.  The phase diagram of the colour multiquarks
nuclear phase, chiral-symmetric~($\chi$S-QGP) phase, and the
vacuum nuclear phase reveals that colour multiquarks are
thermodynamically stable in the region where the temperature is
not too high and the density is sufficiently large~(Figure 8 of
Ref.~\cite{bch}).

In certain situations such as in the core of the neutron stars or
other enormously densed astrophysical objects, exceptionally
strong magnetic field is produced in addition to the high
temperature and density.  Under these fierce conditions, nuclear
matters are pressed together so tightly that deconfinement phase
transition could occur.  As is shown in the phase diagram of
Ref.~\cite{bch}, coloured multiquark states can exist in the
intermediate range of temperature and sufficiently high baryon
chemical potential~(implying high baryon density).  They are
thermodynamically preferred over the other phases such as the
vacuum and the chiral-symmetric deconfined phase of quark-gluon
plasma~($\chi$S-QGP).  It is therefore interesting to explore
magnetic properties of the nuclear phase where coloured exotic
multiquarks exist under these extreme situations. It is possible
that certain classes of densed stars are in the range of
temperature and density suitable for the coloured multiquarks in
the gluon-deconfined soup and the magnetic properties of these
states thus significantly determine their stellar structures.

Responses of the holographic nuclear matter to the external
magnetic field have been intensively investigated in
Ref.~\cite{BLLm0}-\cite{ll}. It was found in Ref.~\cite{BLLm0}
that the external magnetic field makes gluon-deconfined vacuum
more stable thermodynamically than the case when there is no
magnetic field, {\it i.e.} the transition temperature into the
chiral-symmetric quark-gluon plasma increases with the magnetic
field and saturates in the limit of an infinite field.  Authors of
Ref.~\cite{ll} found a phase transition induced by the external
magnetic field in the $\chi$S-QGP phase.  This could be traced
back to the nonlinearity of the DBI action used to describe the
holographic nuclear matter.  Since this transition occurs when the
magnetic field changes from small to large strength, the
Yang-Mills approximation approach~\cite{ts} is no longer valid and
similar transition is not found without consideration of the full
DBI action.  We take the full DBI approach and investigate the
magnetic responses of the multiquark nuclear phase with broken
chiral symmetry in this article.  We found that the magnetized
multiquarks phase are always thermodynamically preferred over the
magnetized vacuum.  At a fixed density, it is also found that the
multiquark states can satisfy the scale fixing condition up to
certain critical values beyond which they would change into
multiquarks with smaller colour charges.  For higher magnetic
fields, all of the multiquarks cannot satisfy the scale fixing
condition at the same density and we would expect other phases to
set in or the density has to be increased for the multiquark
configuration to be able to satisfy the scale fixing condition.

There are two multiquark configurations found below a critical
field. The two configurations merge into one at the critical field
and temperature for a fixed density.  By comparing to the pure
pion gradient and the $\chi$S-QGP phase, the multiquark phase is
found to be preferred thermodynamically at large densities and
moderate fields.

In Sec. 2, the essential features of the multiquarks are reviewed.
Magnetic responses and relevant magnetic phases of the colour
multiquarks are studied in Sec. 3 using the DBI action. Comparison
to the pure pion gradient and the $\chi$S-QGP phase is discussed
in Sec. 4.  We discuss the results and make some conclusions in
Sec. 5.

\section{Exotic multiquark states in the Sakai-Sugimoto model}

In the Sakai-Sugimoto model, gluon deconfinement and the
chiral-symmetry restoration are two distinct phase transitions.
Generically they occur at different temperatures. When the gluons
become deconfined at the deconfinement phase transition, quarks
could still be bound together by the free gluons due to the fact
that the coupling is still strong~(provided that the density is
sufficiently high) and therefore the chiral symmetry could still
be broken.  Due to the deconfinement, the bound states of
multiquarks are not colour singlet in general. Certain properties
of the coloured multiquarks are studied in Ref.~\cite{bch} where
it is demonstrated that the coloured states could exist with
thermodynamical stability. When the temperature rises further, the
bound states become less and less stable and finally completely
dissolved into the quark-gluon plasma.  The chiral symmetry is
restored and everything becomes completely deconfined.

It was proposed by Witten~\cite{witb}, Gross and Ooguri~\cite{go}
that a D-brane wrapping internal subspace of a holographic
background could describe a colour-singlet bound state of $N$
quarks in the dual $U(N)$ gauge theory.  A wrapping D-brane
sources $U(1)$ gauge field on its world volume and induces an $N$
units of $U(1)$ charge upon itself.  This charge needs to be
cancelled by $N$ external strings connecting to the wrapping
brane.  The wrapping brane with $N$ strings attached is called a
baryon vertex.

In the gluon-deconfined phase, more strings can be attached to the
baryon vertex provided that there are equal number of strings
stretching out and go to the background horizon.  This
configuration still conserves the $U(1)$ charge of the brane and
solve the equation of motion of the Nambu-Goto action~\cite{abl}.
We can parameterize the number of radial strings stretching from
the vertex to the horizon as $k_{r}$ and the number of strings
connecting the vertex to the boundary of the background as
$k_{h}$.  For the $k>N$-baryon, $k_{h}-k_{r}=N$ whilst for
$k<N$-baryon, $k_{h}+k_{r}=N$.  Other classes of exotic multiquark
states can be constructed by adding more strings in and out of the
vertex.  Few examples are given in Ref.~\cite{bch} where some
interesting properties are also discussed.

\begin{figure}
\centering
\includegraphics[width=0.8\textwidth]{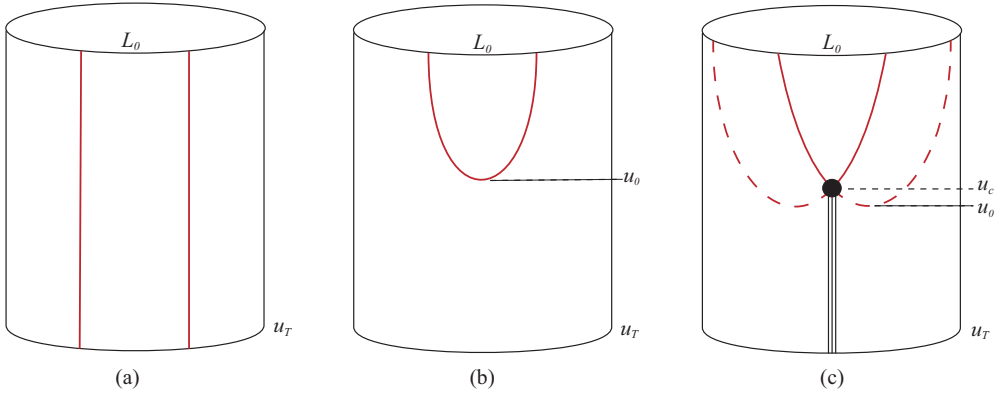}
\caption{Configurations of $\chi$S-QGP~(separate D8,
$\overline{\text{D8}}$)(a), vacuum~(merging D8 and
$\overline{\text{D8}}$)(b) and exotic nuclear phase~(vertex
attached to the D8-$\overline{\text{D8}}$ with radial strings
stretch down to horizon)(c).} \label{config}
\end{figure}

There could exist an interaction among the multiquarks in the form
of connecting strings between each vertex very similar to the
string connecting two end points of quark and antiquark in the
holographic meson configuration.  A multiquark can use one of the
radial strings to merge with another radial string from
neighbouring multiquark and form a colour binding interaction
~(while keeping $k_{h}$ fixed).
 Therefore the number of radial strings represents the {\it colour
charges} of the multiquark.  When the gluons are deconfined, the
``direct" colour interaction would be approximately the same as
the meson and baryon binding potential of the Coulomb type plus
some screening effect. Neglecting the direct interaction and
considering only the DBI-induced collective behaviour of the gas
of multiquarks~\cite{bll,bch}, an approximate phase diagram can be
obtained showing exotic nuclear phase where multiquarks can exist
with thermodynamic stability. Schematic configurations of the
three gluon-deconfined phases are given in Fig.~\ref{config} where
the direction along the circle is the compactified coordinate
$x_{4}$ and the vertical direction is the radial coordinate $u$.

\section{Magnetic properties of the coloured multiquarks in the nuclear phase}

The setup we use is the Sakai-Sugimoto~(SS) model with the source
terms from the instanton embedded in the $D8-\overline{D8}$
configuration, and the radial strings similar to the configuration
used in Ref.~\cite{bch}.  The instanton~(the baryon vertex being
pulled up all the way to the position of the D8-branes by the
strings connecting between the vertex and the flavour branes) is
embedded within the D8-branes and acts as a source for the baryon
density, $d$. The radial strings stretching from the instanton
down to the horizon of the background act as another source.  The
number of radial strings is parameterized by $n_{s}=(\text{number
of radial strings})/N$.  It also tells us how much colour charges
a multiquark has.

The baryon chemical potential is also generated on the D8-branes
by the vector part, $a^{V}_{0}$, of the $U(1)$ subgroup of the
$U(n_{f})$ flavour group of the D8-branes.
 The magnetic field is then turned on by another part of the $U(1)$.  We choose the
direction of the magnetic field so that the vector potential is
\begin{eqnarray}
a^{V}_{3}& = & B x_{2}.
\end{eqnarray}
The vector part $a^{V}_{0}$ is related to the baryon chemical
potential $\mu$ by
\begin{eqnarray}
\mu & = & a^{V}_{0}(u\to\infty),  \nonumber \\
a^{V}_{0}(u_{c}) & = & \mu_{source}, \nonumber \\
\mu_{source} & = & \frac{1}{\mathcal{N}}\frac{\partial
S_{source}}{\partial d}.
\end{eqnarray}
The contributions from the sources, $\mu_{source}$, are from the
baryon vertex and the radial strings.  The full expressions are
given in the Appendix.  The contribution from the $U(1)$ vector
gauge field in the D8-branes, $\mu$, corresponds to the baryon
chemical potential from the content of the plasma.
 The five-dimensional Chern-Simon term of the D8-branes generates
another axial part of the $U(1)$, $a^{A}_{1}$, by coupling it with
$B$ and $a^{V}_{0}$. In this way, the external magnetic field
induces the axial current $j_{A}$ associated with the axial field
$a^{A}_{1}$.

The background metric of the Sakai-Sugimoto model is
\begin{equation}
ds^2=\left( \frac{u}{R_{D4}}\right)^{3/2}\left( f(u) dt^2 +
\delta_{ij} dx^{i}
dx^{j}+{dx_4}^2\right)+\left(\frac{R_{D4}}{u}\right)^{3/2}\left(u^2
d\Omega_4^2 + \frac{du^2}{f(u)}\right)\\ \nonumber
\end{equation}

\begin{equation}
F_{(4)}=\frac{2\pi N}{V_4} {\epsilon}_4, \quad \quad e^{\phi}=g_s
\left( \frac{u}{R_{D4}}\right)^{3/4} ,\quad\quad R_{D4}^3\equiv
\pi g_s N l_{s}^3,\nonumber
\end{equation}

\noindent where $f(u)\equiv 1-u_{T}^{3}/u^3$, $u_T=16{\pi}^2
R_{\text{D4}}^3 {T^2} /9$.  The volume of the unit four-sphere
$\Omega_4$ is denoted by $V_4$ and the corresponding volume 4-form
by $\epsilon_4$.  $l_{s}$ and $g_{s}$ are the string length scale
and the string coupling.  The $x_{4}$ coordinate is compactified
with radius $R$ which is generically different from the curvature
$R_{D4}$ of the background.

The DBI and the Chern-Simon actions of the D8-branes in this
background can be computed to be
\begin{eqnarray}
S_{D8}& = & \mathcal{N}
\int^{\infty}_{u_{c}}du~u^{5/2}\sqrt{1+\frac{B^{2}}{u^{3}}}\sqrt{1+f(u)(a_{1}^{\prime
A})^{2}-(a_{0}^{\prime V})^{2}+f(u)u^{3}x_{4}^{\prime 2}} \\
S_{CS}& = & -\frac{3}{2}\mathcal{N}
\int^{\infty}_{u_{c}}du~(\partial_{2}a^{V}_{3}a^{V}_{0}a^{A
\prime}_{1}-\partial_{2}a^{V}_{3}a^{V \prime}_{0}a^{A}_{1}).
\end{eqnarray}
The normalization factor, $\mathcal{N}=NR^{2}_{D4}/(6\pi^2(2\pi
\alpha^{\prime})^{3})$, represents the brane tension.  The
explanation of the factor $3/2$ is given in Ref.~\cite{BLLm} where
it could be understood as representing the edge effect of the
finite region with uniform magnetic field.
 The equations of motion with respect to $a_{0}^{V},a_{1}^{A}$ are
\begin{eqnarray}
\frac{\sqrt{u^{5}+B^{2}u^{2}}~f(u)a_{1}^{\prime
A}}{\sqrt{1+f(u)(a_{1}^{\prime A})^{2}-(a_{0}^{\prime
V})^{2}+f(u)u^{3}x_{4}^{\prime 2}}}& = & j_{A}-\frac{3}{2}B\mu+3B
a_{0}^{V}, \label{eq:a0} \\
\frac{\sqrt{u^{5}+B^{2}u^{2}}~a_{0}^{\prime
V}}{\sqrt{1+f(u)(a_{1}^{\prime A})^{2}-(a_{0}^{\prime
V})^{2}+f(u)u^{3}x_{4}^{\prime 2}}}& = &
d-\frac{3}{2}Ba_{1}^{A}(\infty)+3B a_{1}^{A}. \label{eq:a1}
\end{eqnarray}
The quantities $d,j_{A}$ are the corresponding density and current
density at the boundary of the background~($u\to\infty$), they are
defined to be
\begin{eqnarray}
j^{\mu}(x, u\to\infty)& = & \frac{\delta S_{eom}}{\delta
A_{\mu}}\bigg{\vert}_{u\to\infty} \\
                      & = & (d,\vec{j_{A}}).
\end{eqnarray}
Explicitly, they are
\begin{eqnarray}
d & = & \frac{\sqrt{u^{5}+B^{2}u^{2}}~a_{0}^{\prime
V}}{\sqrt{1+f(u)(a_{1}^{\prime A})^{2}-(a_{0}^{\prime
V})^{2}+f(u)u^{3}x_{4}^{\prime 2}}}\bigg{\vert}_{\infty}-\frac{3}{2}B a_{1}^{A}(\infty), \\
j_{A}& = & \frac{\sqrt{u^{5}+B^{2}u^{2}}~f(u)a_{1}^{\prime
A}}{\sqrt{1+f(u)(a_{1}^{\prime A})^{2}-(a_{0}^{\prime
V})^{2}+f(u)u^{3}x_{4}^{\prime
2}}}\bigg{\vert}_{\infty}-\frac{3}{2}B\mu.
\end{eqnarray}
For our multiquark configuration, the D8-branes starts from
$u=u_{c}$ and extends to $u\to \infty$.  At the boundary~($u\to
\infty$), the chiral symmetry is broken and therefore the value of
$a^{A}_{1}(\infty)$ is taken to be a physical field,
$\mathcal{5}\varphi$~\cite{BLLm}, describing the degree of chiral
symmetry breaking.  The total action is minimized with respect to
$a^{A}_{1}(\infty)$ if the axial current $j_{A}$~(also defined at
the boundary) is zero.

The total action does not depend on $x_{4}(u)$ explicitly,
therefore the constant of motion leads to
\begin{eqnarray}
(x^{\prime}_{4}(u))^{2}& = & \frac{1}{u^{3}f(u)}\Big[
\frac{u^{3}[f(u)(C(u)+D(u)^{2})-(j_{A}-\frac{3}{2}B\mu
+3Ba_{0}^{V})^{2}]}{F^{2}}-1 \Big]^{-1},  \label{eq:x4prime}
\end{eqnarray}
where
\begin{eqnarray}
F & = & \frac{u^{3}_{c}
\sqrt{f(u_{c})}\sqrt{f(u_{c})(C(u_{c})+D(u_{c})^{2})-(j_{A}-\frac{3}{2}B\mu
+3Ba_{0}^{V}(u_{c}))^{2}}~x_{4}^{\prime}(u_{c})}{\sqrt{1+f(u_{c})u^{3}_{c}~x_{4}^{\prime
2}(u_{c})}}  \label{eq:F}
\end{eqnarray}
and $C(u)\equiv u^{5}+B^{2}u^{2},D(u)\equiv
d+3Ba_{1}^{A}(u)-3B\mathcal{5}\varphi/2$. The expression of
$x_{4}^{\prime}(u_{c})$ is given in the Appendix. It is determined
from the force condition and the scale fixing condition
\begin{equation}
L_{0} = 2 \int^{\infty}_{u_{c}}x_{4}^{\prime}(u)~du = 1.
\label{eq:sf}
\end{equation}

\begin{figure}[htp]
\centering
\includegraphics[width=0.45\textwidth]{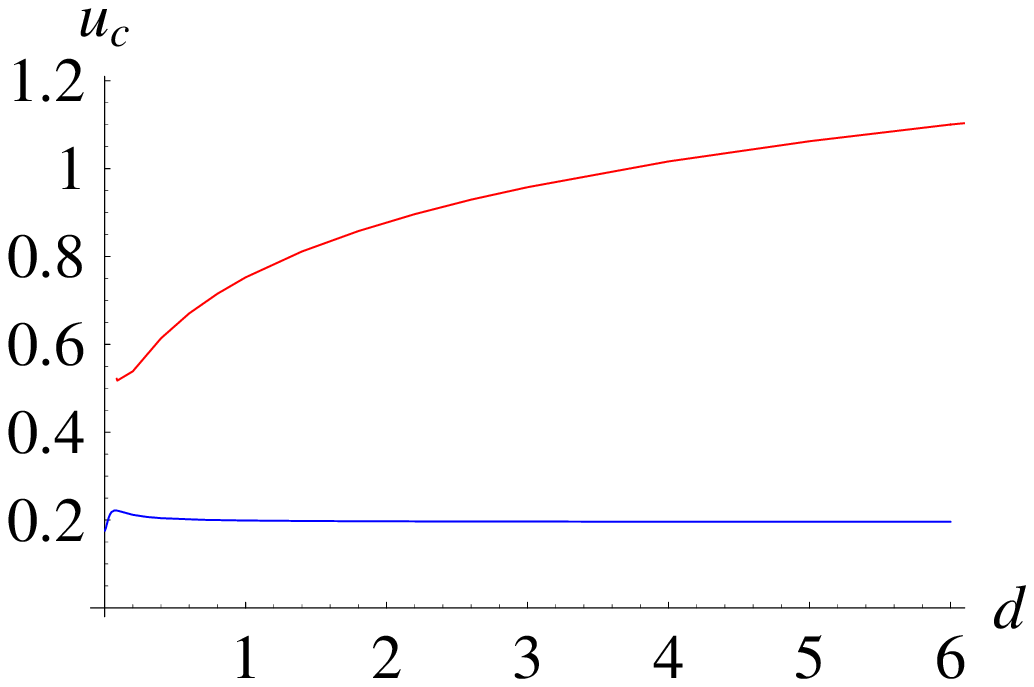} \hfill
\includegraphics[width=0.45\textwidth]{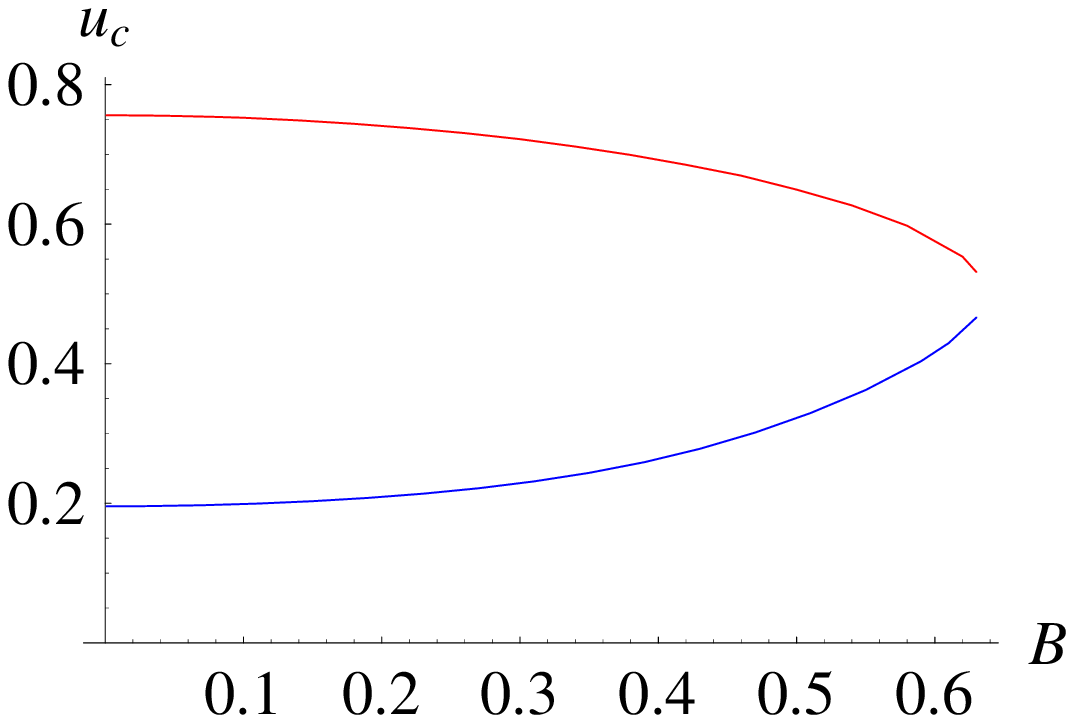}\\
\includegraphics[width=0.45\textwidth]{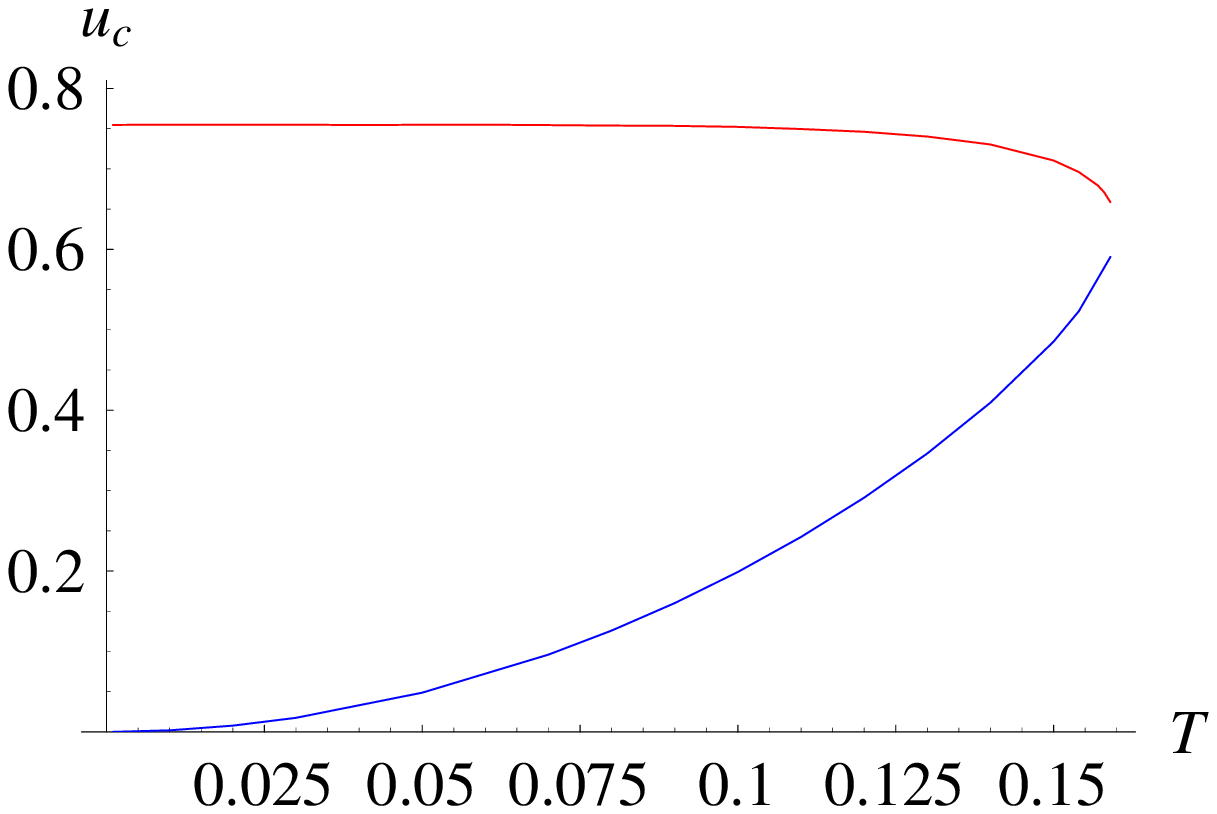}\\
\caption[uc]{Position $u_{c}$ of the vertex for $n_{s}=0$~(normal
baryon) and fixed $j_{A}=0$ as a function of (a) $d$ with fixed
$B=0.10, T=0.10$,(b) $B$ with fixed $d=1, T=0.10$,(c) $T$ with
fixed $B=0.10, d=1$. The lower~(blue) line is the
configuration-\textbf{A} with $u_{c}$ close to $u_{T}$ and the
upper~(red) line is the configuration-\textbf{B} with large
separation between $u_{c}$ and $u_{T}$.} \label{figuc}
\end{figure}

Since $x_{4}^{\prime}(u)$ depends on both
$a_{0}^{V}(u),a_{1}^{A}(u)$, we need to solve the differential
equations (\ref{eq:a0}) and (\ref{eq:a1}) with $x_{4}^{\prime}(u)$
substituted into the equations of motion and check whether the
solutions satisfy the scale fixing condition Eqn.~(\ref{eq:sf}).
The values of the vector and axial field at the vertex are also
chosen so that $a^{V}_{0}(u_{c})=\mu_{source},a^{A}_{1}(u_{c})=0$.
We basically perform the shooting algorithm by choosing the value
of $\mu$ and $\mathcal{5}\varphi$ in the expression for
$x_{4}^{\prime}(u_{c})$ until we hit $a^{V}_{0}(\infty)=\mu$ and
$a^{A}_{1}(\infty)=\mathcal{5}\varphi$. If the resulted solution
satisfies the scale fixing condition $L_{0}=1$, we keep the
solution.  If not, we adjust the value of $u_{c}$ and perform the
shooting procedure again.  The position $u_{c}$ for $n_{s}=0$ is
given as a function of the density, the magnetic field, and the
temperature in Fig.~\ref{figuc}.

\begin{figure}[htp]
\centering
\includegraphics[width=0.45\textwidth]{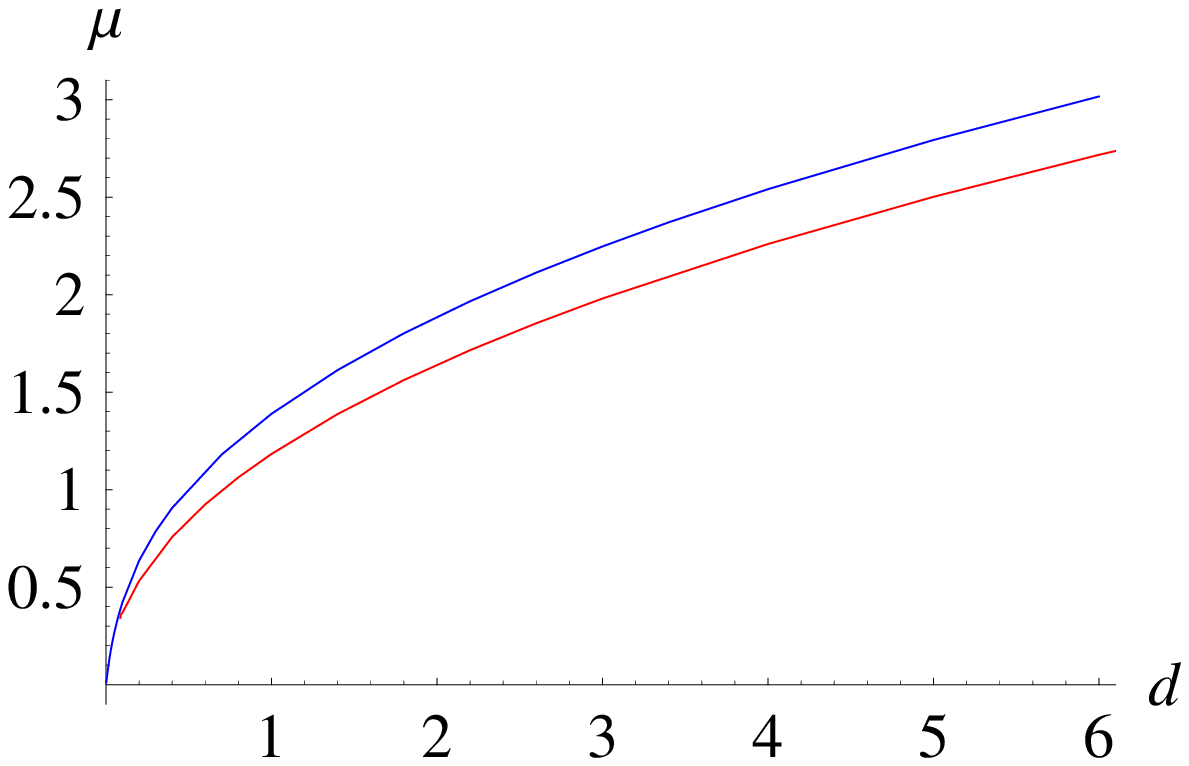} \hfill
\includegraphics[width=0.45\textwidth]{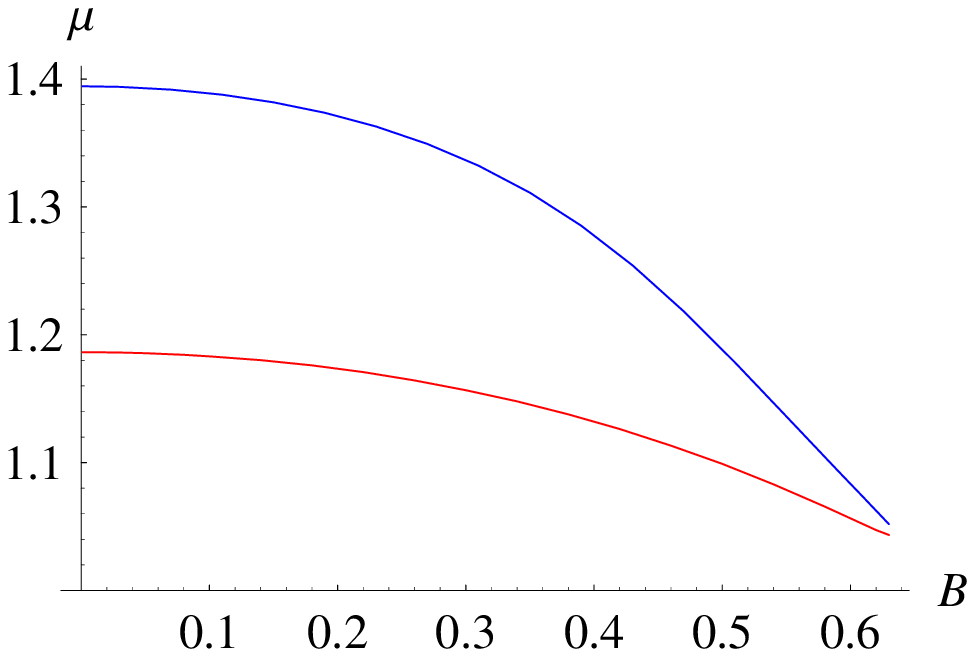}\\
\includegraphics[width=0.45\textwidth]{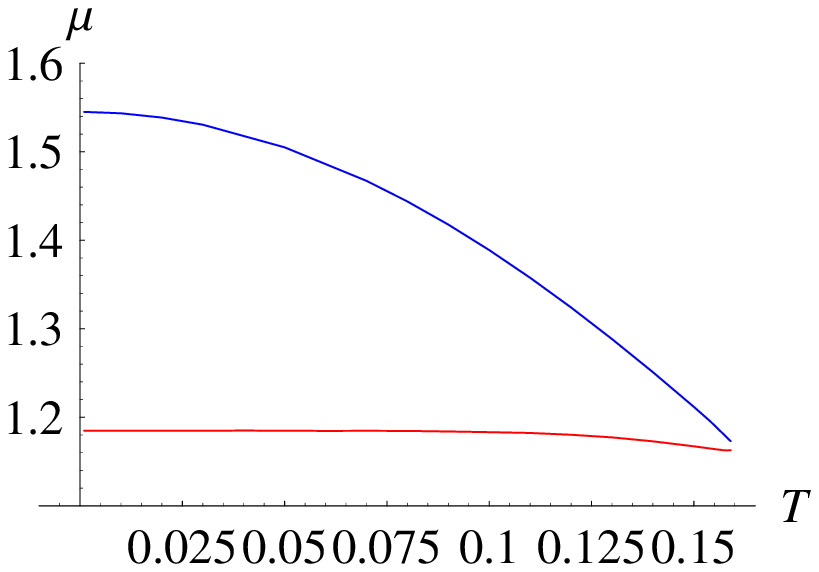}\\
\caption[baryon chemical potential]{The baryon chemical potential
$\mu$ for $n_{s}=0$~(normal baryon) and fixed $j_{A}=0$ as a
function of (a) $d$ with fixed $B=0.10, T=0.10$,(b) $B$ with fixed
$d=1, T=0.10$,(c) $T$ with fixed $B=0.10, d=1$.  The upper~(blue)
line is the configuration-\textbf{A} with $u_{c}$ close to $u_{T}$
and the lower~(red) line is the configuration-\textbf{B} with
large separation between $u_{c}$ and $u_{T}$.} \label{fig1}
\end{figure}

\begin{figure}[htp]
\centering
\includegraphics[width=0.45\textwidth]{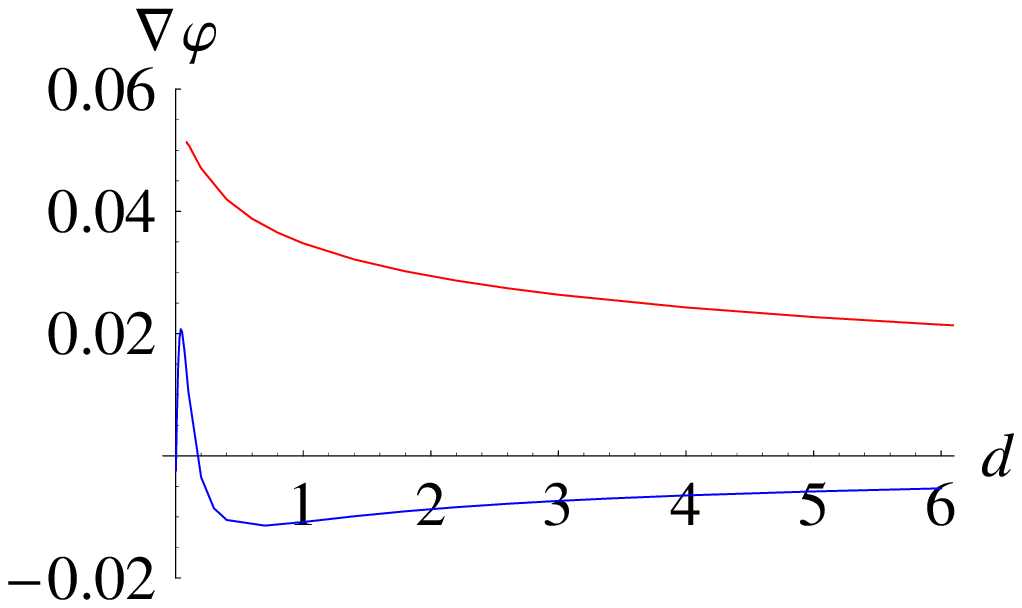} \hfill
\includegraphics[width=0.45\textwidth]{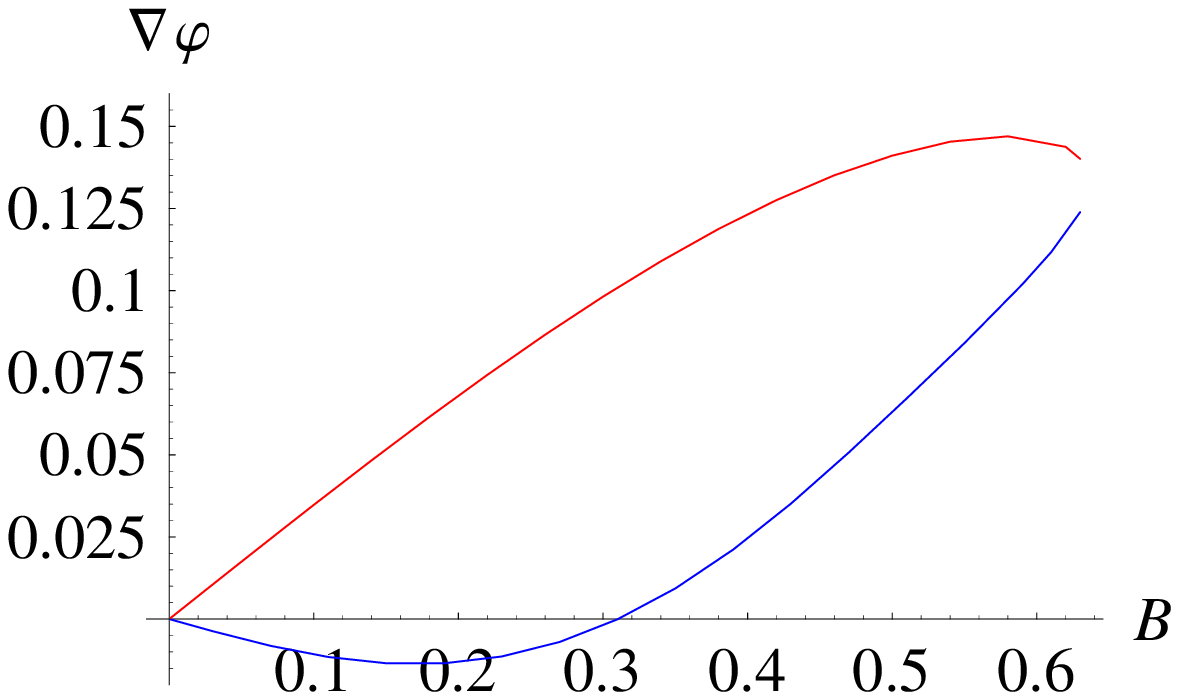}\\
\includegraphics[width=0.45\textwidth]{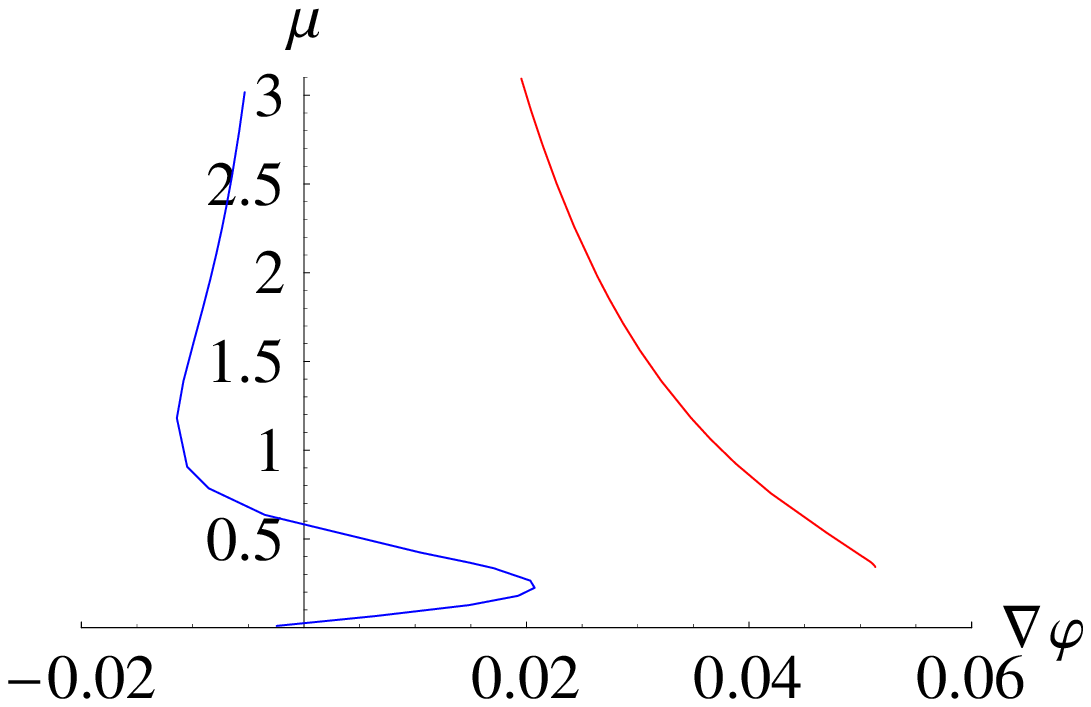} \hfill
\includegraphics[width=0.45\textwidth]{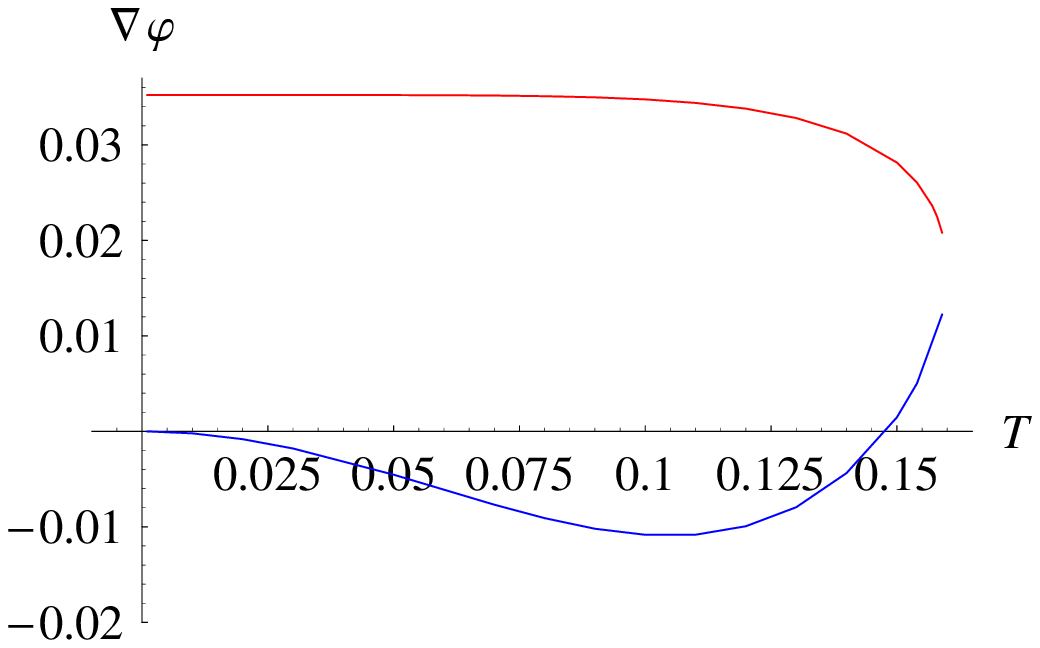}\\
\caption[baryon chemical potential]{The gradient of the scalar
field $\mathcal{5}\varphi \equiv a^{A}_{1}(\infty)$ for
$n_{s}=0$~(normal baryon) and fixed $j_{A}=0$ as a function of (a)
$d$ with fixed $B=0.10, T=0.10$,(b) $B$ with fixed $d=1,
T=0.10$,(c) $\mu$ with fixed $B=0.10, T=0.10$,(d) $T$ with fixed
$B=0.10, d=1$.  The lower~(blue) line is the
configuration-\textbf{A} with $u_{c}$ close to $u_{T}$ and the
upper~(red) line is the configuration-\textbf{B} with large
separation between $u_{c}$ and $u_{T}$.} \label{fig1.1}
\end{figure}

From the solutions of the equations of motion, the relations
between baryon chemical potential~($\mu$) and the baryon
density~($d$), the magnetic field~($B$), and the temperature~($T$)
are obtained for the choice of parameters $n_{s}=0$~(normal
baryon), $j_{A}=0$, as are shown in Figure \ref{fig1}.  There are
two types of solution corresponding to the two holographic
multiquark configurations.  One is the configuration with $u_{c}$
close to $u_{T}$~(configuration-\textbf{A}) and another is the
configuration with a large separation between $u_{c}$ and
$u_{T}$~(configuration-\textbf{B}).

The baryon chemical potential is found to be an increasing
function of the density for most range of $d$ for both
configuration \textbf{A,B}.  As is shown in Fig.~\ref{figuc},
configuration-\textbf{A} has the position of vertex $u_{c}$ closer
to the horizon $u_{T}$ than configuration-\textbf{B}.  At very
small $d$, the two configurations emerge separately as two
distinct configurations.  Interestingly, as the magnetic field and
temperature increase, the two configurations converge into a
single configuration as we can see the position $u_{c}$ approaches
the same value at the critical field and temperature~(see
Fig.~\ref{figuc}).  However, when the two configurations merge,
the configuration no longer satisfies the scale fixing condition
$L_{0}=1$ and we expect it to change into other phases such as the
chiral-symmetric quark-gluon plasma for a fixed density.  It turns
out that if the density is allowed to change, the multiquark
configuration can continue to satisfy the scale fixing condition
at higher fields provided that the density is sufficiently large.
 This will be discussed more in Section \ref{compare}.

In Fig.~\ref{fig1}, the baryon chemical potential is an increasing
function of $d$, this is true for both configuration-\textbf{A}
and \textbf{B}. It is roughly a linear function of the density,
showing that the DBI-induced collective interaction between the
multiquarks are negligible.  As $d$ gets larger, the DBI-induced
effect sets in and the negative binding interaction makes $\mu$
grows with $d$ less quickly than the linear progression. Note that
this DBI-induced interaction occurs even when the baryon is colour
singlet due to the nonlinear nature of the DBI action. The origin
of this DBI-induced interaction is the ``{\it tidal weight}" of
the DBI action contributed from both the branes' worldsheet metric
and the background gauge field strength. Naturally, any form of
energy contributes to the tidal weight even the colour singlets.

For configuration-\textbf{B}, there seems to be minimal density
$d_{min}$ below which the shooting algorithm could not find other
valid solutions.  We are not certain what happens below these
values.  It is possible that when the field is turned on, the
D8-branes acquire higher tension and therefore the configuration
requires minimal density to pull it down in order for the distance
between D8 and $\overline{\text{D8}}$ to reach $L_{0}=1$.  For
$T=0.10, B=0.10, n_{s}=0$, the value of $d_{min}$ for multiquark
configuration \textbf{B} is approximately 0.086.

Figure~\ref{fig1} shows that the chemical potential is a
decreasing function with respect to the magnetic field.  This is
similar to the behaviour of baryons in chiral-symmetric
quark-gluon plasma studied in Ref.~\cite{BLLm}.  When the field
gets stronger up to certain values, the field becomes too strong
for the force condition to hold at the scale fixing $L_{0}=1$.
This strange behaviour is shown in Fig.~\ref{fig2} where
multiquarks with smaller $n_{s}$ are shown to be able to exist up
to stronger fields.

As is also shown in Fig.~\ref{fig1}, the relationship between
$\mu$ and $T$ is as we expect, a decreasing function of $T$ for
fixed density $d$ since higher temperature will melt the
multiquarks away. For fixed $d$ and $B$, the multiquark
configuration satisfies the scale fixing condition up to a maximum
temperature above which we expect it to melt into the plasma.  For
$n_{s}=0$, this critical temperature is about $0.159$ for $d=1$.

It is interesting to note that the baryon chemical potential of
$n_{s}=0$ multiquarks for both configurations converge to the same
value at critical field~($\simeq 0.63$) and temperature~($\simeq
0.159$) for $d=1$. This behaviour also shows up in the gradient
scalar field as is shown in Fig.~\ref{fig1.1}.

Fig.~\ref{fig1.1} shows the relations between the field
$\mathcal{5}\varphi$ and the density, the magnetic field, the
baryon chemical potential, and the temperature.  The pion gradient
$\mathcal{5}\varphi$ represents the domain wall of the scalar
field induced by the magnetic field on the nuclear
vacuum~\cite{sst}.  Roughly speaking, it quantifies the degree of
chiral symmetry breaking.  The domain wall carries baryon charge
and thus contributes to the baryon density.  For multiquark
configuration-\textbf{B}, it increases with $B$ for a fixed
density. From Fig.~\ref{fig1.1}, the pion gradient field increases
linearly with respect to the field for small fields.  Then it
starts to saturate closed to the critical field.  This is somewhat
similar to the behaviour of the pion gradient in the confined
phase studied in Ref.~\cite{BLLm}.  For configuration-\textbf{B},
the pion gradient field is a decreasing function of the density
when the field is fixed.  This implies that for a fixed magnetic
field, the population of the domain wall becomes lesser as the
density of the baryon~(including multiquarks and other bound
states) increases.  We will see this behaviour again in Sect. 4
when we consider the pure pion gradient phase.  Finally from
Fig.~\ref{fig1.1}, the degree of chiral symmetry breaking
$\mathcal{5}\varphi$ decreases as temperature rises for multiquark
configuration-\textbf{B}.

For configuration-\textbf{A}, the pion gradient field decreases at
first for small magnetic fields, but turns to rise with the field
around $B\approx 0.16$ until it converges to
configuration-\textbf{B} at the critical field.  The dependence of
the field $\mathcal{5}\varphi$ on the density at a fixed
$B=0.10,T=0.10$ shows a minimum at $d\approx 0.7$, corresponding
to $\mu \approx 1.18$.  Then as the density grows, the pion
gradient increases and saturates, implying limited contribution of
the domain wall for large baryon density.  The temperature
dependence of the pion gradient field for multiquark
configuration-\textbf{A} shows some peculiar behaviour.  First, it
becomes more negative at low temperatures then turns to rise and
converge to configuration-\textbf{B} at the critical temperature.

Figure \ref{fig2.1} shows how the pion gradient field
$\mathcal{5}\varphi$ varies with the magnetic field $B$ for
$n_{s}=0, 0.10, 0.20$.  For the same $B$, multiquarks with higher
$n_{s}$ responds more to the magnetic field by inducing larger
$\mathcal{5}\varphi$, implying higher degree of chiral symmetry
breaking.  The pion gradient field for both multiquark
configuration-\textbf{A,B} forms a butterfly-wing shape graph for
each $n_{s}$.  The edge of the wing is at the critical field where
the configuration converges and barely satisfies the scale fixing
condition.

\begin{figure}[htp]
\centering
\includegraphics[width=0.45\textwidth]{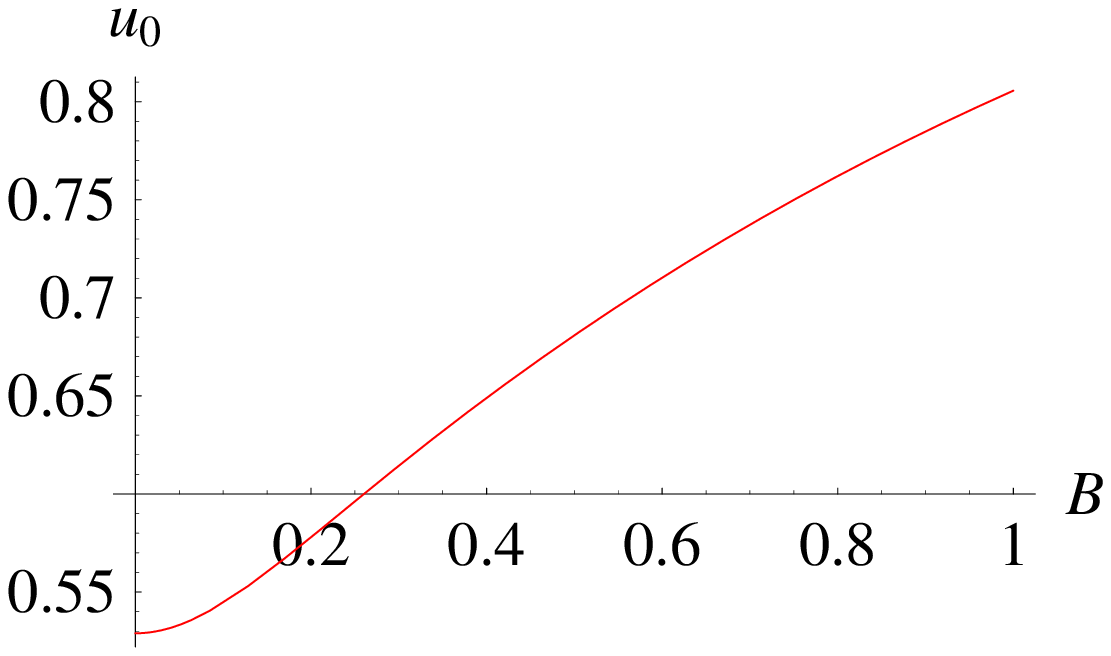} \hfill
\includegraphics[width=0.45\textwidth]{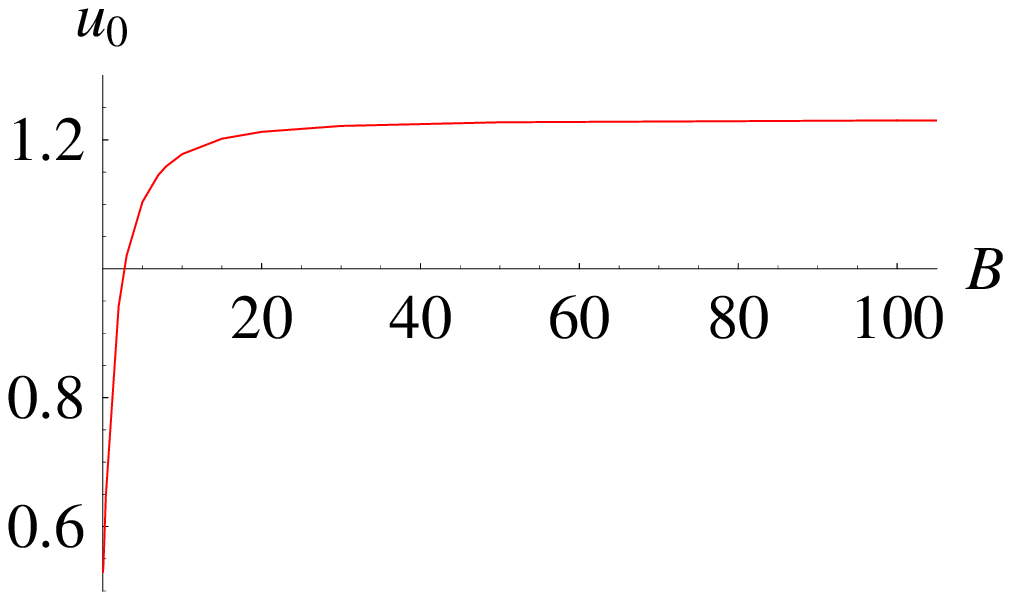}\\
\caption[u0B relation]{Relation between $u_{0}$ and external
magnetic field $B$ of the vacuum for the temperature $T=0.10$,
$u_{0}$ saturates to the approximate value of $1.23$ at large
field.}\label{fig1.5}
\end{figure}

\begin{figure}[htp]
\centering
\includegraphics[width=0.6\textwidth]{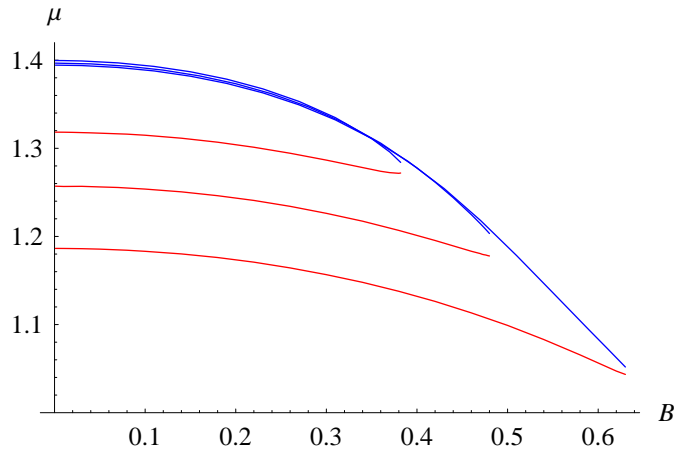}
\caption[muB relation]{Comparison between the baryon chemical
potential as a function of $B$ at fixed $j_{A}=0, d=1, T=0.10$ and
(a) $n_{s}=0$~(normal baryon), the bottom graph,(b) $n_{s}=0.10$,
the middle graph, (c) $n_{s}=0.20$, the top graph.  The
upper~(blue) lines are the configuration-\textbf{A} with $u_{c}$
close to $u_{T}$ and the lower~(red) lines are the
configuration-\textbf{B} with large separation between $u_{c}$ and
$u_{T}$.}\label{fig2}
\end{figure}

\begin{figure}[htp]
\centering
\includegraphics[width=0.6\textwidth]{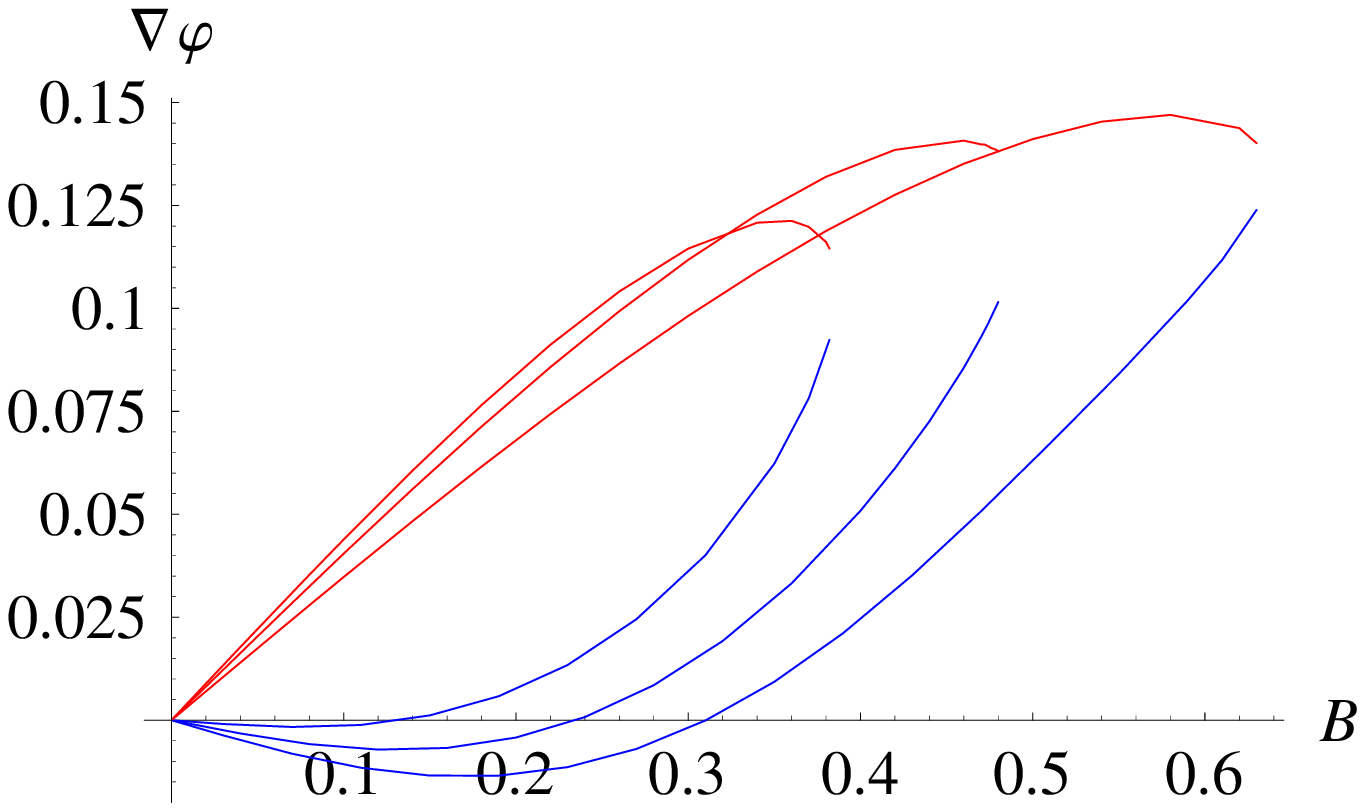}
\caption[yB relation]{Comparison between the gradient of the
scalar field $\mathcal{5}\varphi$ as a function of $B$ at fixed
$j_{A}=0, d=1, T=0.10$ and (a) $n_{s}=0$~(normal baryon), the
bottom graph,(b) $n_{s}=0.10$, the middle graph, (c) $n_{s}=0.20$,
the top graph.  The lower~(blue) lines are the
configuration-\textbf{A} with $u_{c}$ close to $u_{T}$ and the
upper~(red) lines are the configuration-\textbf{B} with large
separation between $u_{c}$ and $u_{T}$.}\label{fig2.1}
\end{figure}

The magnetization of the multiquarks nuclear matter can be defined
using the regulated free energy,
$\mathcal{F}_{\text{E}}=\Omega(\mu,B) +\mu d$, in the canonical
ensemble as
\begin{eqnarray}
M(d,B)& = & -\frac{\partial \mathcal{F}_{\text{E}}(d,B)}{\partial
B}\bigg{\vert}_{d},
\end{eqnarray}
where $\Omega(\mu,B) =
S[a_{0}(u),a_{1}(u)](e.o.m.)-S[\text{magnetized vacuum}]$.  The
action with $a_{0}^{\prime V}, a_{1}^{\prime A}$ eliminated is
given by $S[a_{0}(u),a_{1}(u)](e.o.m.)= S_{D8}+S_{CS}$ where
\begin{equation}
S_{D8}=\mathcal{N}
\int^{\infty}_{u_{c}}du~C(u)\sqrt{\frac{f(u)(1+f(u)u^{3}{x^{\prime
2}_{4})}}{f(u)(C(u)+D(u)^{2})-(j_{A}-\frac{3}{2}B\mu
+3Ba_{0}^{V})^{2}}}, \nonumber
\end{equation}
and $S_{CS}$ is given in the Appendix.  The grand canonical
potential is regulated with respect to the magnetized vacuum.  The
action of the magnetized vacuum with non-vanishing
$x^{\prime}_{4}$ is
\begin{eqnarray}
S[\text{magnetized vacuum}] & = &
\int^{\infty}_{u_{0}}~\sqrt{C(u)(1+f(u)u^{3}x^{\prime
2}_{4})}\bigg{\vert}_{vac}~du, \nonumber
\end{eqnarray}
where
\begin{eqnarray}
 x^{\prime}_{4}(u)\vert_{vac} & = &
\frac{1}{\sqrt{f(u)u^{3}\Big(\frac{f(u)u^{3}C(u)}{f(u_{0})u^{3}_{0}C(u_{0})}-1
\Big)}}.
\end{eqnarray}
The position $u_{0}$ where $x^{\prime}_{4}(u_{0})=\infty$ of the
magnetized vacuum can be solved numerically from $L_{0}=1$~(with
$u_{0}$ replacing $u_{c}$ in the limit of integration). The
relation between $u_{0}$ and the magnetic field is shown in
Fig.~\ref{fig1.5} for $T=0.10$. As the magnetic field gets
stronger, the position of the lowest position of the
D8-$\overline{\text{D8}}$ configuration, $u_{0}$, becomes larger,
in order to satisfy the condition $L_{0}=1$~(implying heavier
branes due to magnetic field energy). At $T=0.10$, position of
$u_{0}$ saturates to the value of about $1.23$~(The number changes
with temperature, of course) in the limit of an infinite field.

\begin{figure}[htp]
\centering
\includegraphics[width=0.6\textwidth]{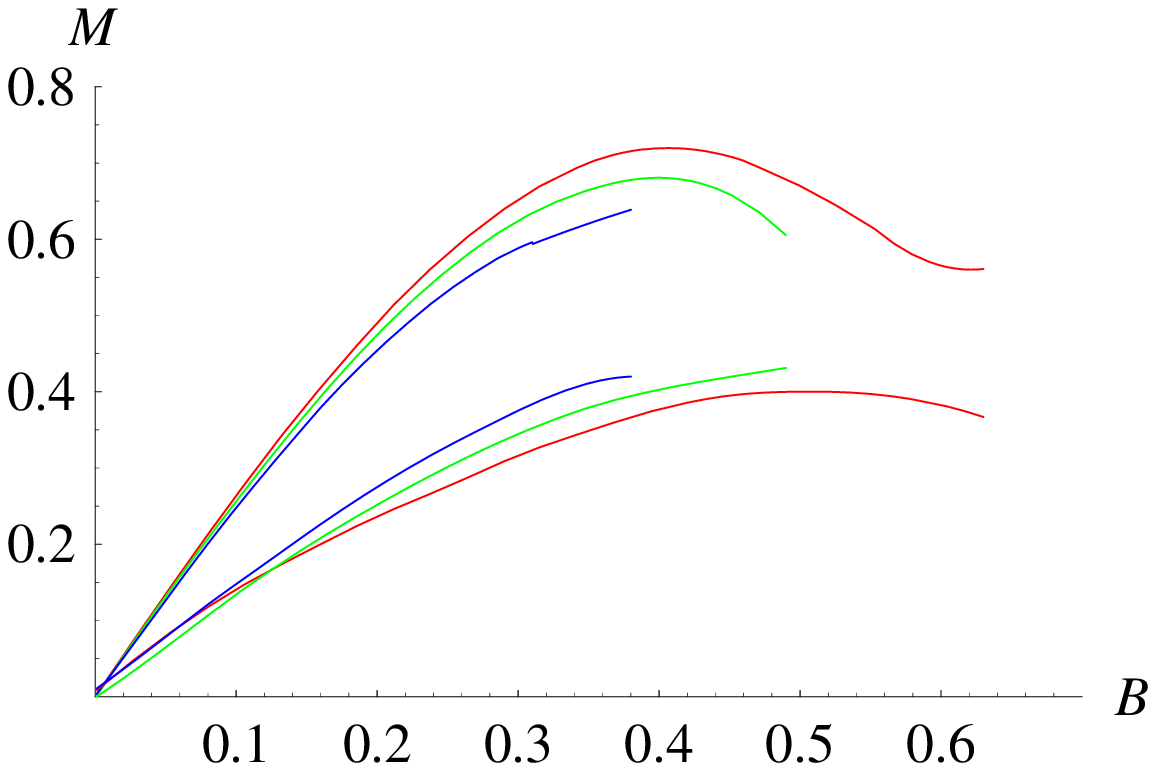}
\caption[magnetization]{The magnetization of the multiquarks
nuclear matter at fixed $j_{A}=0, d=1$, and $T=0.10$ for
$n_{s}=0$~(red), $0.10$~(green), $0.20$~(blue).  The upper lines
are the configuration-\textbf{A} with $u_{c}$ close to $u_{T}$ and
the lower lines are the configuration-\textbf{B} with large
separation between $u_{c}$ and $u_{T}$.}\label{fig3}
\end{figure}

The magnetization of the multiquark nuclear matter is shown in
Fig.~\ref{fig3} for $n_{s}=0$~(red), $0.10$~(green),
$0.20$~(blue). The magnetization is positive and increases as $B$
increases until the field is close to the critical value then it
starts to drop.  Generically, configuration-\textbf{A} of
multiquarks has larger magnetization than
configuration-\textbf{B}.  For the configuration-\textbf{B~(A)},
multiquarks with higher~(lower) $n_{s}$ have higher
magnetizations.  As the magnetic field gets stronger beyond the
critical field for each $n_{s}$, the multiquarks will undergo a
transition into the ones with smaller $n_{s}$.  For even larger
fields, even the $n_{s}=0$ multiquarks cannot satisfy the scale
fixing condition if the density is not allowed to change.

\begin{figure}[htp]
\centering
\includegraphics[width=0.6\textwidth]{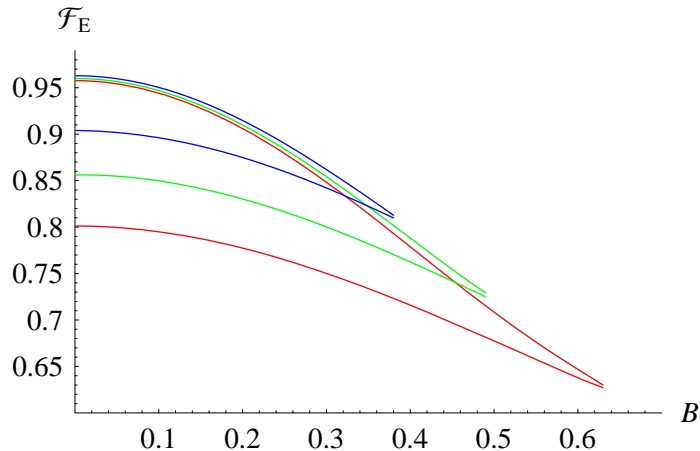}
\caption[freeE]{The free energy of the multiquarks nuclear matter
at fixed $j_{A}=0, d=1$, and $T=0.10$ for $n_{s}=0$~(red),
$0.10$~(green), $0.20$~(blue).  The upper lines are the
configuration-\textbf{A} with $u_{c}$ close to $u_{T}$ and the
lower lines are the configuration-\textbf{B} with large separation
between $u_{c}$ and $u_{T}$.}\label{fig4}
\end{figure}

Interestingly, numerical studies reveal that the grand canonical
potential of the multiquark phase is always lower than the grand
canonical potential of the magnetized vacuum, i.e.
$S[a_{0}(u),a_{1}(u)](e.o.m.)-S[\text{magnetized vacuum}]<0$, for
the entire range of $B$.  This suggests that once $\mu >
\mu_{onset}$, the magnetized multiquark phase is always
thermodynamically preferred over the magnetized vacuum, the
situation similar to the case when there is no magnetic field
investigated in Ref.~\cite{bch}.  Among the two configurations, we
found from Fig.~\ref{fig4} that the free energy of
configuration-\textbf{B} is always lower than
configuration-\textbf{A} and thus more stable thermodynamically.
These two multiquark configurations-\textbf{A,B} are the long and
short cusp configurations discussed in Ref.~\cite{bll}, being
extended to the general case with nonzero magnetic fields. It is
found here that for a fixed density, strong field and/or high
temperature~(see Fig.~\ref{fig1} and \ref{fig1.1}) converge the
two into a single configuration right before dissociating them
altogether.

Figure~\ref{fig4} shows how the free energy changes with the
magnetic field for $n_{s}=0$~(red), $0.10$~(green), $0.20$~(blue)
at the temperature $T=0.10$ and the density $d=1$.  For each
$n_{s}$, both configurations converge to the same
configuration~(with the same baryon chemical potential, degree of
chiral symmetry breaking and free energy) at the critical fields.
The critical fields for $n_{s}=0, 0.1, 0.2$ are roughly $0.63,
0.48, 0.38$ respectively.

\section{Comparison to other phases} \label{compare}

In this section we compare the baryon chemical potential and free
energy of the magnetized multiquarks to the pure pion gradient
phase and the chiral symmetric quark-gluon plasma~($\chi_S$-QGP)
phase, both under the external magnetic field with gluons
deconfined. The pure pion gradient phase is defined to be the
phase with $\mu_{source}=0$~(sourceless case) and the baryon
chemical potential comes purely from the induced gradient field,
$\mathcal{5}\varphi$, in response to the external field.  The
baryon density also comes purely from the pion gradient
field~($d=3B\mathcal{5}\varphi/2$). A similar situation in the
confined phase of the antipodal SS model has been studied in
Ref.~\cite{BLLm}. The $\chi_S$-QGP under the presence of the
external magnetic field has been explored in Ref.~\cite{BLLm,ll}
but again only limited to the antipodal case of the SS model.  In
this section we explore some of their magnetic properties in more
general case where $x^{\prime}_{4}(u)$ is not zero and the scale
is fixed to $L_{0}=1$.  Even though the extra constraints are
irrelevant to the $\chi_S$-QGP~(since $x^{\prime}_{4}=0$ for this
configuration), it makes crucial difference in the case of pure
pion gradient phase.  The scale fixing condition is found to be
very difficult for the pure pion gradient configuration to satisfy
for most of the density as we will discuss below.

All three phases under consideration obey the same set of
equations of motion, Eqn.~(\ref{eq:a0}),(\ref{eq:a1}) with each
specific set of the
boundary conditions and parameters as the following,  \\

\underline{multiquark phase}: $j_{A}=0,
\mu_{source}=a^{V}_{0}(u_{c}), \mathcal{5}\varphi =
a^{A}_{1}(\infty), a^{A}_{1}(u_{c})=0$, \\

\underline{pure pion gradient phase}: all the same with the
multiquark phase with the following exceptions,
 $\mu_{source}=0,a^{V}_{0}(u_{c})\neq 0,d=\frac{3}{2}B \mathcal{5}\varphi$, \\

\underline{$\chi_S$-QGP}: $x^{\prime}_{4}(u)=0$ and
$\mathcal{5}\varphi = a^{A}_{1}(\infty)=0,
\mu_{source}=a^{V}_{0}(u_{c}=u_{T})=0,j_{A}=\frac{3}{2}B\mu$~(since
the configuration extends to $u_{T}$ and $f(u_{T})=0$ so that Eqn.~(\ref{eq:a0}) is zero). \\

First, we will explore certain properties of the pure pion
gradient phase and show that it does not exist in the range of
parameters~($d\geq 1, B\leq 1-2$) under consideration.  Then
comparison between the multiquark and the $\chi_S$-QGP phases will
be discussed.

\subsection{Pure pion gradient phase}

\begin{figure}[htp]
\centering
\includegraphics[width=0.45\textwidth]{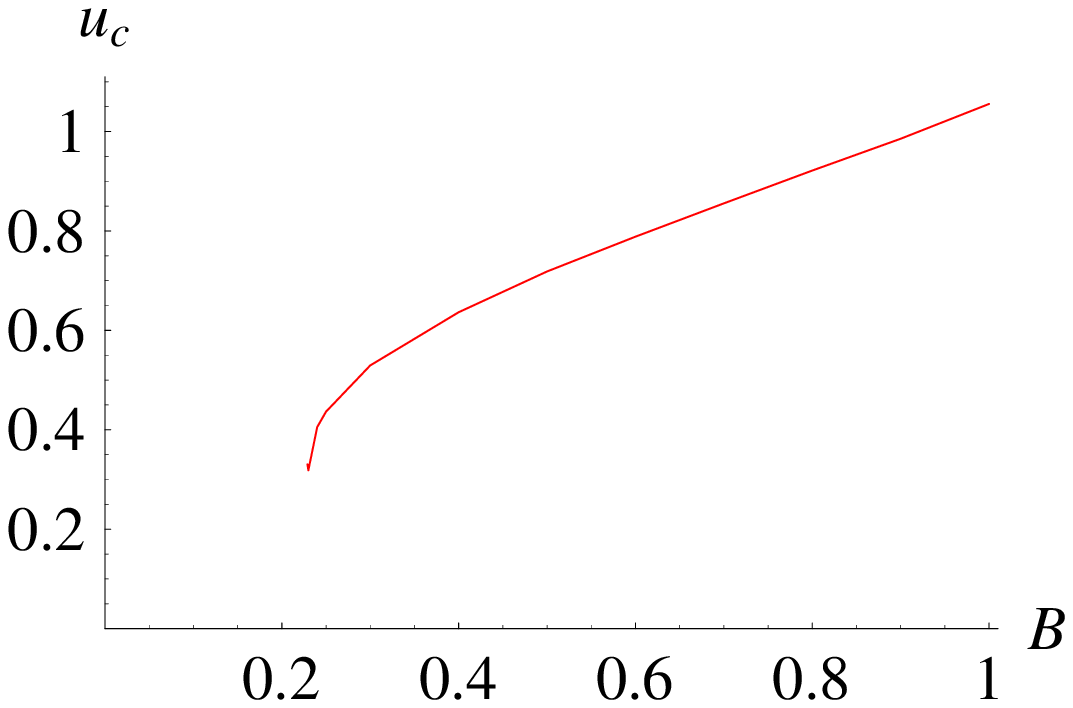} \hfill
\includegraphics[width=0.45\textwidth]{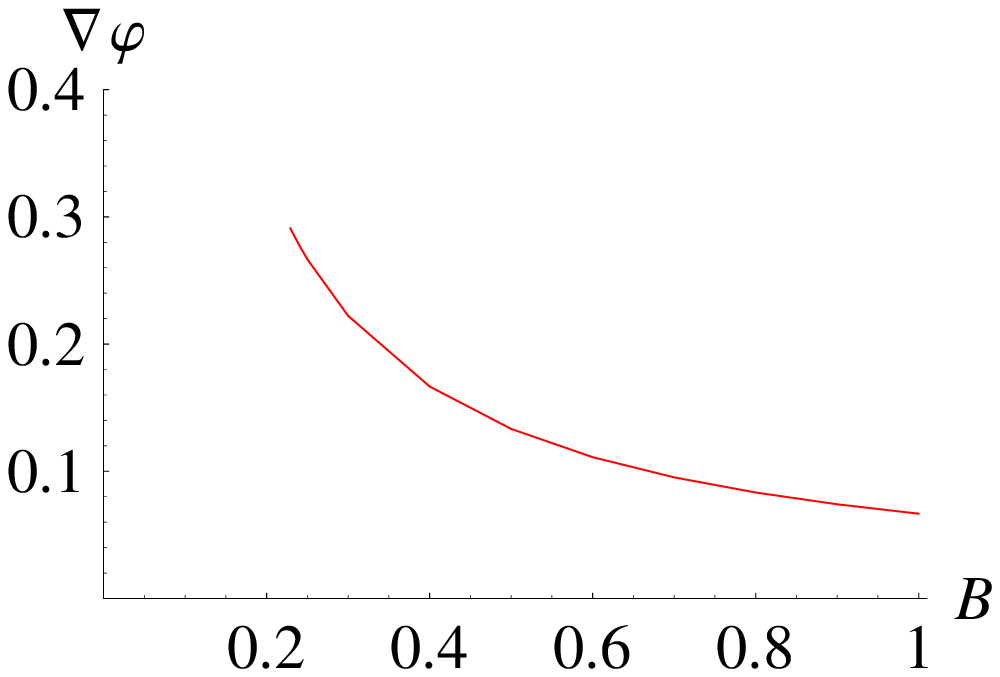}\\
\includegraphics[width=0.45\textwidth]{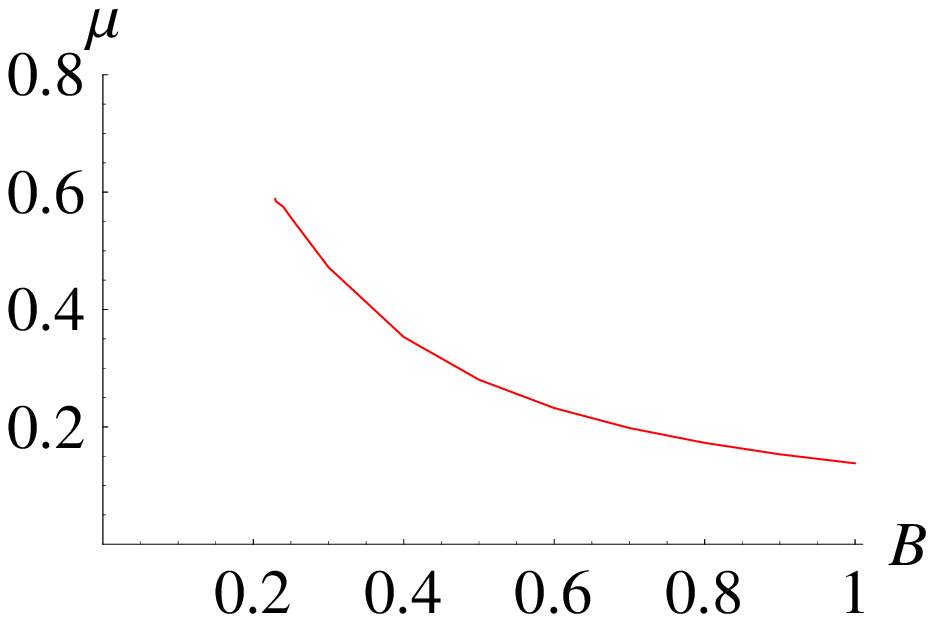}\\
\caption[Pure Pion phase]{The position $u_{c}$, the pion gradient
field, and the baryon chemical potential of the pure pion gradient
phase at $d=0.10,T=0.10$ as a function of the magnetic field.}
\label{figPi}
\end{figure}

For pure pion gradient configuration, the contribution of the
sources, the vertex and strings, is set to zero..  Effectively, we
set $\mu_{source}=0, d=3B\mathcal{5}\varphi/2$.  This is because
when $\mathcal{5}\varphi$ is zero, the density $d$ should
represent the density of the sources, i.e. the pure multiquark or
pure baryon configuration, therefore the source density should be
given by $d-\frac{3}{2}B\mathcal{5}\varphi$ on the right-hand side
of Eqn.~(\ref{eq:a1}).  When we fix the value of the density at a
fixed magnetic field, $\mathcal{5}\varphi$ is also fixed.  For
example, when $d=1, B=0.1$, $\mathcal{5}\varphi \simeq 6.667$, a
relatively large value.  This large value of
$\mathcal{5}\varphi=a^{A}_{1}(\infty)$ leads to a generically
large value of $a^{A}_{1}(u)$ for the most range of $u$.  From
Eqn.~(\ref{eq:x4prime}) and (\ref{eq:F}), we see that for the pure
pion phase, $D(u)=3B a^{A}_{1}(u)$ and thus it must be large for
the most range of $u$ as well.  In the multiquark configuration,
the $d$ dependence of $D(u_{c})$ in the expression of $F$,
Eqn.~(\ref{eq:F}), will compensate the largeness of $D(u)$ and
$x^{\prime}_{4}$ can be made sufficiently large so that $L_{0}=1$
could still be satisfied.  However, in pure pion phase, $D(u_{c})$
is simply zero.  This makes $x^{\prime}_{4}$ getting smaller as
the density gets larger and the scale fixing condition $L_{0}=1$
would not be satisfied above certain value of the density for a
fixed $B$.

As a result, we wish to keep $\mathcal{5}\varphi$ sufficiently
small in order to satisfy the scale fixing condition.  This
implies that higher densities require larger magnetic fields.  To
demonstrate this, we fix baryon density to $d=0.1$ and plot the
position $u_{c}$ of the vertex and the baryon chemical potential
as a function of the magnetic field in Fig.~\ref{figPi}.  The
graph of $u_{c}$ shows a minimal field at about $B\approx 0.229$
below which $L_{0}<1$ for all solutions.  For a larger density
$d\geq 1$, the required field strengths are $B>>1$ in order for
the scale fixing condition to be satisfied.  For the range of
parameters $d=1.0, B \leq 1.0$, we therefore need to consider only
the two phases of the multiquark and the $\chi_S$-QGP.  The same
situation occurs for the range of parameters $d=10,B \leq 1-2$
where the pure pion gradient phase does NOT satisfy the scale
fixing condition and therefore does not exist as well.

\subsection{Multiquark-domain wall versus $\chi_S$-QGP phase}

\begin{figure}[htp]
\centering
\includegraphics[width=0.6\textwidth]{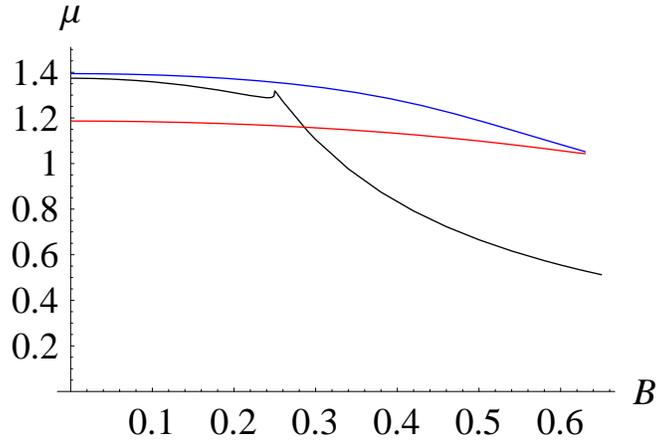}
\caption[chemical potential comparison]{Comparison between the
baryon chemical potential for $T=0.10$ at a fixed density $d=1$ of
(a) $n_{s}=0$~(normal baryon) multiquark configuration-\textbf{A},
the top~(blue) graph,(b) $\chi_S$-QGP, the middle~(black)
graph,(c) $n_{s}=0$~(normal baryon) multiquark
configuration-\textbf{B}, the bottom~(red) graph.
}\label{figmucom}
\end{figure}

\begin{figure}[htp]
\centering
\includegraphics[width=0.6\textwidth]{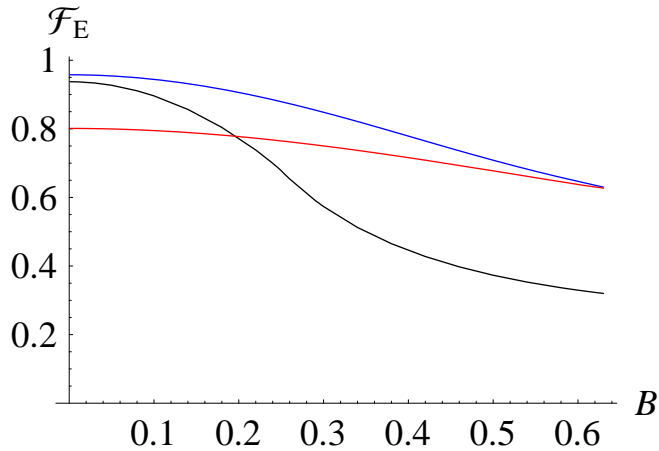}
\caption[free energy comparison]{Comparison between the free
energy for $T=0.10$ at a fixed density $d=1$ of (a)
$n_{s}=0$~(normal baryon) multiquark configuration-\textbf{A}, the
top~(blue) graph,(b) $\chi_S$-QGP, the middle~(black) graph,(c)
$n_{s}=0$~(normal baryon) multiquark configuration-\textbf{B}, the
bottom~(red) graph . }\label{figFcom}
\end{figure}

\begin{figure}[htp]
\centering
\includegraphics[width=0.6\textwidth]{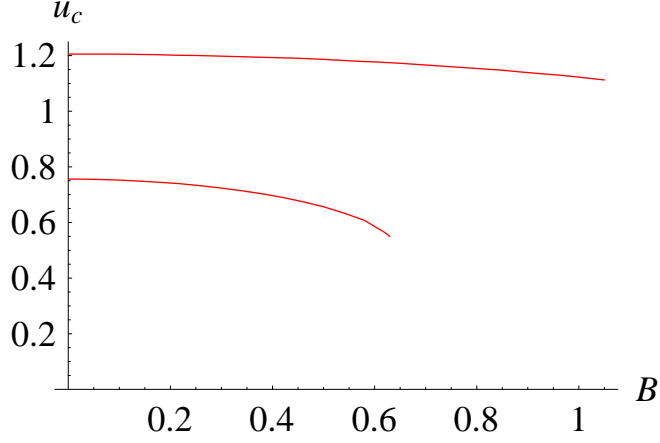}
\caption[free energy comparison]{The position of the vertex
$u_{c}$ as a function of $B$ at a fixed density $d=1$~(lower) and
$d=10$~(upper) for $T=0.10$ of $n_{s}=0$~(normal baryon)
multiquark configuration-\textbf{B} phase.}\label{figubBD10}
\end{figure}

\begin{figure}[htp]
\centering
\includegraphics[width=0.6\textwidth]{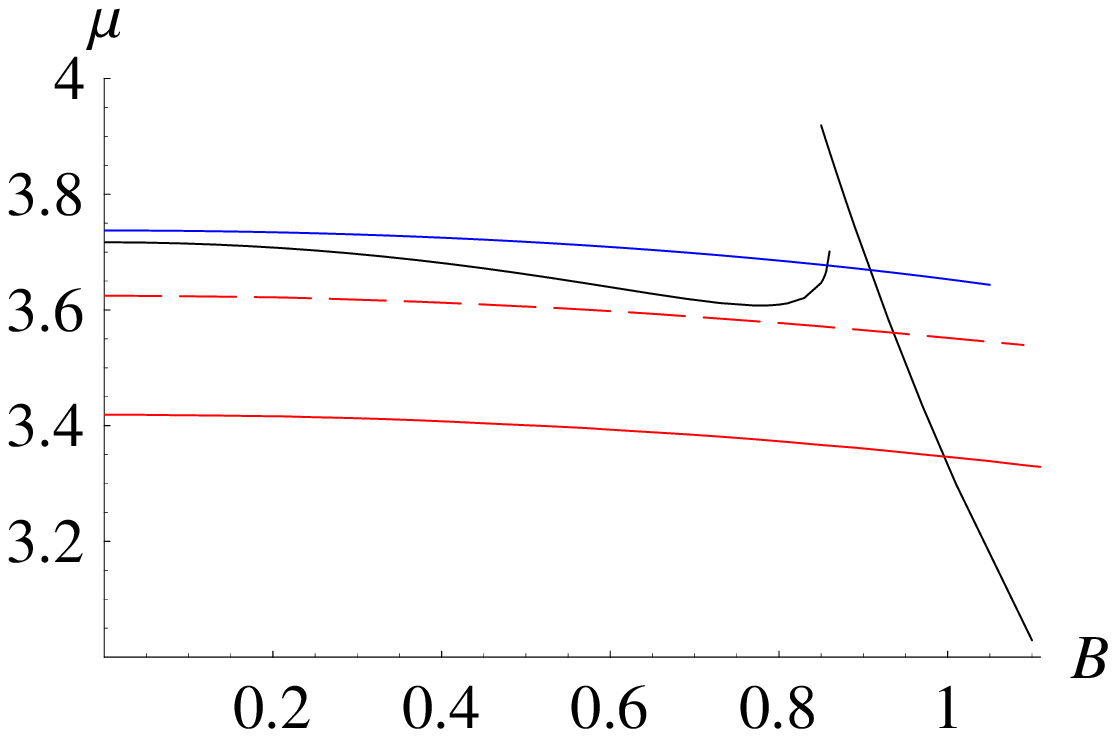}
\caption[mu comparison]{Comparison between the baryon chemical
potential for $T=0.10$ at a fixed density $d=10$ of (a)
$n_{s}=0$~(normal baryon) multiquark configuration-\textbf{A}, the
top~(blue) curve,(b) $\chi_S$-QGP, the black curve,(c) $n_{s}=0.2$
multiquark configuration-\textbf{B}, the dashed red curve,(d)
$n_{s}=0$~(normal baryon) multiquark configuration-\textbf{B}, the
red curve. }\label{figmucomD10}
\end{figure}

\begin{figure}[htp]
\centering
\includegraphics[width=0.6\textwidth]{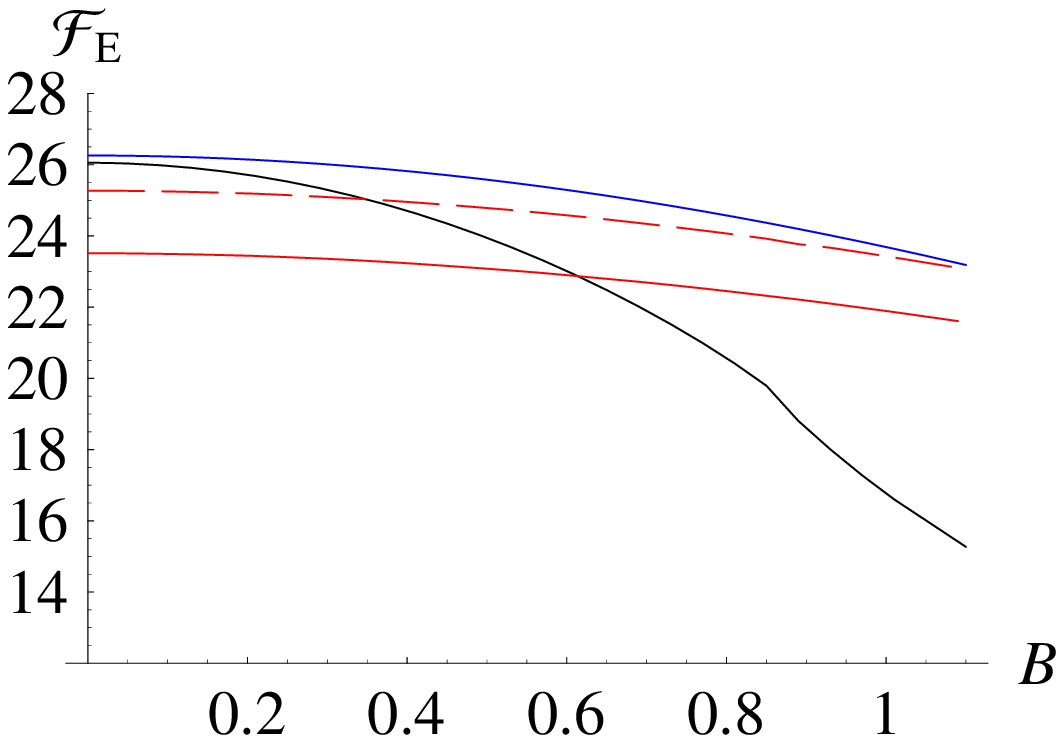}
\caption[free energy comparison]{Comparison between the free
energy for $T=0.10$ at a fixed density $d=10$ of (a)
$n_{s}=0$~(normal baryon) multiquark configuration-\textbf{A}, the
top~(blue) curve,(b) $\chi_S$-QGP, the black curve,(c) $n_{s}=0.2$
multiquark configuration-\textbf{B}, the dashed red curve,(d)
$n_{s}=0$~(normal baryon) multiquark configuration-\textbf{B}, the
red curve. }\label{figFcom10}
\end{figure}

\begin{figure}[htp]
\centering
\includegraphics[width=0.6\textwidth]{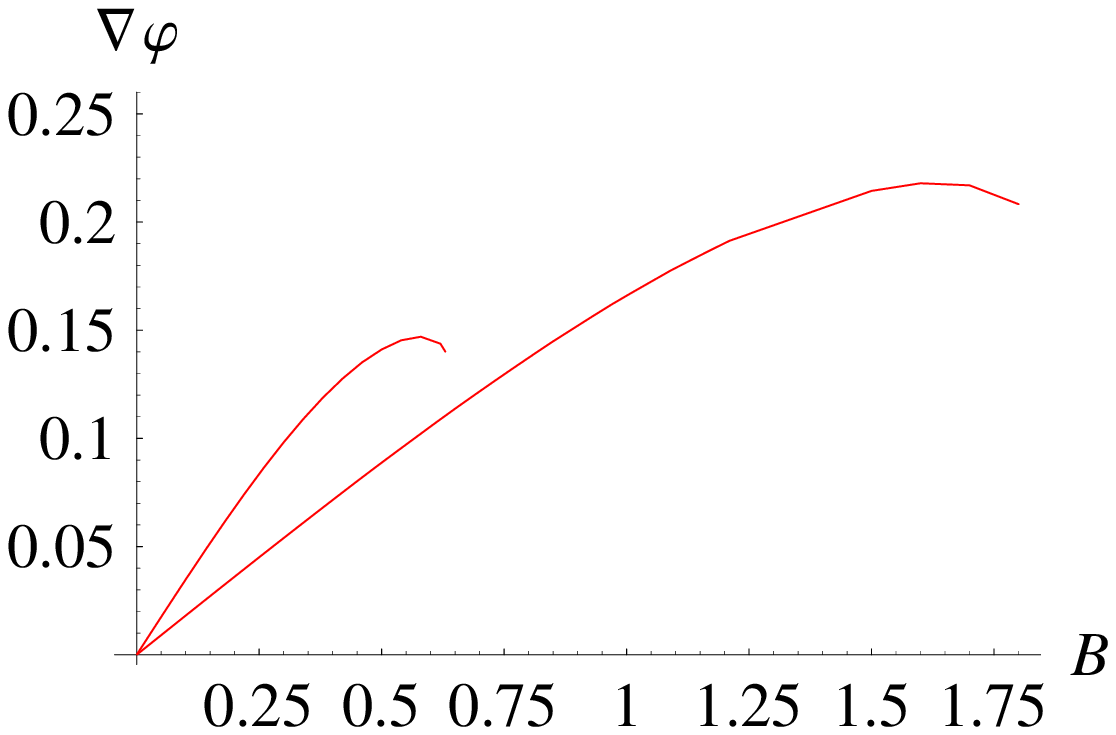}
\caption[Pion and B]{Plots between the pion gradient field of the
multiquark phase and the magnetic field for $T=0.10,n_{s}=0$ at
$d=1$~(shorter) and $d=10$~(longer). }\label{figyBD10}
\end{figure}

The baryon chemical potential $\mu$ is to be found by shooting
algorithm for a fixed $d,B,T$, for each phase.  For
$d=1,B=0.10,T=0.10$, they are shown in Fig.~\ref{figmucom}.
Observe that there are two possible solutions for the $\chi_S$-QGP
phase. As the magnetic field increases beyond a certain value~(in
this case around $B\approx 0.25$), there will be phase transition
to another solution within this phase.  This behaviour is explored
in details in Ref.~\cite{ll}.  When the density is raised to
$d=10$, the transition occurs at higher field around $B\approx
0.86$~(Fig.~\ref{figmucomD10}).  The transitions can also be seen
in the plots of the free energy,
Fig.~\ref{figFcom},\ref{figFcom10}, where the slopes of the graphs
change abruptly around the critical fields.  For $d=1$, this is
quite small and somewhat hard to see but it becomes apparent for
$d=10$.

From the plots of the free energy, Fig.~\ref{figFcom}, the
multiquark configuration-\textbf{A} is the least preferred phase
when the density is small~($d=1$).  Its free energy is larger than
the $\chi_S$-QGP phase for all fields.  For $B\leq 0.196$, the
most preferred phase is the multiquark configuration-\textbf{B}
phase with the lower free energy.  The $\chi_S$-QGP phase is more
stable for $d=1, B>0.196$.  Nevertheless, the multiquark
configurations can exist up to only about the critical fields
beyond which they cannot satisfy the scale fixing condition at
that particular density.

However, this does not mean that the multiquarks phase cannot
exist in the range of field larger than the critical value.
Stronger field gives the D8-branes larger tension and thus it
requires sufficiently heavier vertex and strings to pull it down
in order for the distance between D8 and $\overline{\text{D8}}$ to
reach $L_{0}=1$. This implies that we need larger $d$ in order to
make the configuration satisfy the scale fixing condition at
stronger fields.
Fig.~\ref{figubBD10},\ref{figmucomD10},\ref{figFcom10} confirm
this insight.  They show the plots of the multiquarks
configurations when the density is large~($d=10$). Multiquark
configurations can exist far beyond the critical field $B\approx
0.63$ of the small $d$ case~($d=1$).  In particular,
Fig.~\ref{figFcom10} demonstrates that at $d=10$, the multiquark
configurations~($n_{s}=0,0.2$), with lower free energies, are
thermodynamically preferred over the $\chi_S$-QGP for $B<0.61$ and
$B<0.348$ respectively.

It is thus reasonable to conclude that for larger densities, the
multiquarks phase will be more and more preferred over the
$\chi_S$-QGP phase, in a larger and larger range of the field.
Magnetized multiquarks and the induced pion gradient field are
thus stable thermodynamically and they will mix together in the
magnetized nuclear~(multiquark-domain wall) phase provided that
the density is sufficiently large and the temperature is not too
high.

Finally for completeness, we present the plots of the pion
gradient field of the multiquark phase~(Fig.~\ref{figyBD10}), the
pion gradient field becomes smaller for a given $B$ as the density
increases. However, it extends to larger range of fields for
larger density.  We can therefore conclude that at the large
densities~(and baryon chemical potential), contribution of the
pion gradient becomes lesser and the multiquarks contribute
dominantly to the baryon density and chemical potential.  This is
also shown in Fig.~\ref{fig1.1}.

\section{Discussions and Conclusion}

In Sakai-Sugimoto model, chiral symmetry restoration and gluon
deconfinement are two distinct phase transitions.  Generically,
with an exception of the antipodal case with $x^{\prime}_{4}=0$,
gluon deconfinement occurs at lower temperature than the chiral
symmetry restoration. For the region of the phase diagram between
the two transitions, coloured multiquarks can exist with
thermodynamical stability~(the phase diagram is shown in Figure 8
of Ref.\cite{bch}).

Magnetic responses of the nuclear phase with colour multiquarks
are studied here by using one component of the $U(1)$ subgroup of
$U(N_{f})$ as the vector potential of the external magnetic field.
The Chern-Simon action of the D8-branes couples the magnetic field
to an axial vector component, $a^{A}_{1}$, of the $U(1)$, inducing
axial current $j_{A}$.  When the chiral symmetry is broken, we
effectively set $j_{A}$ to zero.  The value of $a^{A}_{1}(\infty)$
then describes the degree of chiral symmetry breaking of the
phase.

There are two possible multiquark configurations \textbf{A} and
\textbf{B}.
 Configuration-\textbf{A} is the configuration where the baryon
vertex is close to the horizon.  Configuration-\textbf{B}, on the
other hand, has the baryon vertex more separated from the horizon.
By comparing the free energy of the two configurations in
Fig.~\ref{fig4}, we found that configuration-\textbf{B} is more
stable themodynamically.
 We establish relations between the baryon chemical potential and
the baryon density, the external magnetic field, and the
temperature for both configurations as are shown in
Fig.~\ref{fig1}. Baryon chemical potential is an increasing
function of the density when the field is turned on. This is the
same behaviour to the case when there is no field.

On the other hand, the relation between chemical potential and the
magnetic field is rather interesting.  The baryon chemical
potential is a decreasing function of the field. For multiquarks
with high value of $n_{s}$~(number of radial strings), the
configuration finds it more difficult to satisfy the scale fixing
condition at large fields.  There is a maximum field strength for
each $n_{s}$ above which the multiquark configuration cannot
exist~(Fig.~\ref{fig2}).  This is in contrast to the behaviour of
the chiral-symmetric quark-gluon plasma~(in the antipodal case of
the Sakai-Sugimoto model with no instantons, {\it i.e.}
$x^{\prime}_{4}(u)=0$ case) studied in Ref.~\cite{BLLm} where
chemical potential is always a decreasing function with respect to
$B$ and the configuration continues to exist at arbitrarily large
fields.  This is due to the fixation of the density. Stronger
field gives the flavour branes more tension and when the field is
too strong, a fixed density source would not be sufficiently heavy
to pull the branes down for the distance between D8 and
$\overline{\text{D8}}$ to reach $L_{0}=1$. Temperature also has
effect on the multiquarks, sufficiently high temperature will melt
away the multiquarks even in the presence of an external field.

The gradient of the scalar field representing the chiral symmetry
breaking, $\mathcal{5}\varphi=a^{A}_{1}(\infty)$, is found to
roughly increase in magnitude with the field.  For the same field
strength and fixed density, multiquarks with higher $n_{s}$~(i.e.
larger colour charges) show higher degree of chiral symmetry
breaking~(larger magnitude of $a^{A}_{1}(\infty)$), but can only
sustain the force condition up to smaller fields as is shown in
Fig.~\ref{fig2.1}.

The mixing of pion gradient with the miltiquark in the multiquark
phase decreases as the density increases~(Fig.~\ref{fig1.1}).  It
is found that the pure pion gradient phase~(no multiquark
contribution) does not satisfy the scale fixing condition for
large densities and moderate fields.

What would happen if the magnetic field increases beyond the point
where the multiquarks can satisfy the scale fixing condition
$L_{0}=1$?  We would expect the multiquarks to change into the
multiquarks with lower $n_{s}$ as is shown in Fig.~\ref{fig2} for
a fixed $d$ and $T$ since they can still satisfy the scale fixing
condition.  This induces a sudden drop in the baryon chemical
potential. Also in the situation where $\mu$ is kept fixed instead
of $d$, the multiquarks are forced to jump to the larger $d$ in
order to change into the multiquarks with lower $n_{s}$ as the
field increases beyond the critical point.  For even larger
fields, all of the multiquarks cannot satisfy the scale fixing
condition for a fixed density.  There would be phase transition to
other phase.  For a fixed density, the phase will change into the
$\chi_S$-QGP.  However, if we allow the density to change~(in a
more realistic situation), the system could change into the
multiquark~(with pion gradient mixing) phase for a sufficiently
large density.  The phase of multiquark with pion gradient mixing
is found to be more preferred than the $\chi_S$-QGP at large
densities~(implying large baryon chemical potentials) and moderate
fields. This is shown in Fig.~\ref{figFcom10}.

For configuration-\textbf{B} multiquarks, The magnetization of the
multiquark nuclear matter is found to be an increasing function of
$B$ for $n_{s}=0, 0.10, 0.20$  except when the fields get close to
the critical points.  Close to the critical fields, the
magnetizations saturate and even start to decrease.  The
magnetized multiquarks phases are thermodynamically preferred over
the magnetized vacuum once the baryon chemical potential is higher
than the onset value~($\mu
> \mu_{onset}$). This is similar to the case when there is no
magnetic field investigated in Ref.~\cite{bch}.

\section*{Acknowledgments}
\indent I would like to thank Ekapong Hirunsirisawat for help with
the first illustration.  P.B. is supported in part by the Thailand
Research Fund~(TRF) and Commission on Higher Education~(CHE) under
grant MRG5180227.

\appendix
\section{Force condition of the multiquark configuration}

The forces on the D4-brane in the flavour D8-branes are balanced
among three forces from the tidal weight of the D4-brane, the
force from the strings attached to the D4, and the force from the
D8-branes.  Varying the total action with respect to $u_{c}$ gives
the surface term.  Together with the scale-fixing condition
$2\int^{\infty}_{u_{c}}~du x^{\prime}_{4}(u)=L_{0}=1$, we
obtain~\cite{bll}
\begin{eqnarray}
x^{\prime}_{4}(u_{c})& = & \displaystyle{ \left( \tilde{L}(u_{c})
-
\frac{\partial{S_{source}}}{\partial{u_{c}}}\right)\Bigg{/}{\frac{\partial
\tilde{S}}{\partial{x^{\prime}_{4}}}\bigg{\vert}_{u_{c}}}},
\label{app1}
\end{eqnarray}
as the condition on $u_{c}$.

The Legendre transformed action is given by
\begin{eqnarray}
\tilde{S}& = &
\int^{\infty}_{u_{c}}\tilde{L}(x^{\prime}_{4}(u),d)\,du, \nonumber \\
              & = &
\mathcal{N}
\int^{\infty}_{u_{c}}du~\sqrt{\frac{1}{f(u)}+u^{3}x_{4}^{\prime
2}} \nonumber
\\
& \times & \sqrt{f(u)(C(u)+D(u)^{2})-\Big(j_{A}-\frac{3}{2}B\mu
+3Ba_{0}^{V}\Big)^{2}},\label{app2}
\end{eqnarray}
where $C(u)\equiv u^{5}+B^{2}u^{2},D(u)\equiv d+3Ba_{1}^{A}(u)-3B
\mathcal{5}\varphi/2$. It is calculated by performing Legendre
transformation with respect to $a^{V\prime}_{0}$ and
$a^{A\prime}_{1}$ respectively. Note that the Chern-Simon action
is also included in the total action during the transformations.

The Chern-Simon term with the derivatives $a^{V \prime},a^{A
\prime}$ eliminated is
\begin{eqnarray}
S_{CS}& = & -\mathcal{N} \frac{3}{2}B
\int^{\infty}_{u_{c}}du~\frac{\Big(
a^{V}_{0}(j_{A}-\frac{3}{2}B\mu +3Ba_{0}^{V})-f(u)
D(u)a^{A}_{1}\Big)\sqrt{\frac{1}{f(u)}+u^{3}x_{4}^{\prime
2}}}{\sqrt{f(u)(C(u)+D(u)^{2})-\Big(j_{A}-\frac{3}{2}B\mu
+3Ba_{0}^{V}\Big)^{2}}}.\label{app3}
\end{eqnarray}

 Lastly, in order to compute $x^{\prime}_{4}(u_{c})$ we consider the source
 term~\cite{bch}
\begin{eqnarray}
S_{source}     & = & {\mathcal N} d \Big[
\frac{1}{3}u_{c}\sqrt{f(u_{c})}+n_{s}(u_{c}-u_{T})\Big],
\\ \label{app4}
               & = & {\mathcal N} d \mu_{source}
\end{eqnarray}
where $n_{s}=k_{r}/N$ is the number of radial strings in the unit
of $1/N$.

From Eqn.~(\ref{app1}),(\ref{app2}),(\ref{app3}),(\ref{app4}), and
setting $a^{V}_{0}(u_{c})=\mu_{source},a^{A}_{1}(u_{c})=0$ we can
solve to obtain
\begin{eqnarray}
(x^{\prime}_{4}(u_{c}))^{2}& = &
\frac{1}{f_{c}u_{c}^{3}}\Big[\frac{9}{d^{2}}\frac{
(f_{c}(C_{c}+D_{c}^{2})-(j_{A}-\frac{3}{2}B\mu+3 B
a_{0}^{V}(u_{c}))^{2})}{(1+\frac{1}{2}(\frac{u_{T}}{u_{c}})^{3}+3
 n_{s}\sqrt{f_{c}})^{2}}
-1\Big] \nonumber
\end{eqnarray}
where $f_{c}\equiv f(u_{c}),C_{c}\equiv C(u_{c}),D_{c}\equiv
D(u_{c})$.

When we fix the parameter $n_{s}$, the temperature $T$, the baryon
density $d$, the axial current $j_{A}=0$~(by minimizing the action
with respect to $a^{A}_{1}(\infty)$), and setting
$a^{A}_{1}(u_{c})=0, a^{V}_{0}(u_{c})=\mu_{source}$, then the
position $u_{c}$ of the D4-brane is completely determined as a
function of the magnetic field $B$. Once the equations of motion
are solved, the value of $\mu=a^{V}_{0}(\infty)$ and
$a^{A}_{1}(\infty)$ are determined.

\end{document}